\documentclass[11pt]{article}%
\usepackage{amsmath}
\usepackage{amsfonts}
\usepackage{amssymb}
\usepackage{a4wide}%
\newtheorem{theorem}{Theorem}[section]
\newtheorem{corollary}[theorem]{Corollary}

\newtheorem{lemma}[theorem]{Lemma}
\newtheorem{proposition}[theorem]{Proposition}
\newtheorem{remark}[theorem]{Remark}
\numberwithin{equation}{section}
\begin{document}

\title{On the entropy minimization problem in Statistical Mechanics}
\author{Constantin Z\u{a}linescu\thanks{Faculty of Mathematics, University Al.\ I.
Cuza, Bd.\ Carol I, Nr. 11, 700506 Ia\c{s}i, Romania, e-mail:
\texttt{zalinesc@uaic.ro}. }}
\date{}
\maketitle

\begin{abstract}
In the works on Statistical Mechanics and Statistical Physics, when deriving
the distribution of particles of ideal gases, one uses the method of Lagrange
multipliers in a formal way. In this paper we treat rigorously this problem
for Bose--Einstein, Fermi--Dirac and Maxwell--Boltzmann entropies and present
a complete study in the case of the Maxwell--Boltzmann entropy. Our approach
is based on recent results on series of convex functions.

\end{abstract}



\section{Introduction}

In Statistical Mechanics and Statistical Physics, when studying the
distribution of the particles of an ideal gas, one considers the problem of
maximizing
\begin{equation}
\sum_{i}\left[  n_{i}\ln\left(  \frac{g_{i}}{n_{i}}-a\right)  -\frac{g_{i}}%
{a}\ln\left(  1-a\frac{n_{i}}{g_{i}}\right)  \right]  \label{PB}%
\end{equation}
with the constraints $\sum_{i}n_{i}=N$ and $\sum_{i}n_{i}\varepsilon_{i}=E,$
where, as mentioned in \cite[pp.\ 141--144]{PatBea:11}, $\varepsilon_{i}$
denote the average energy of a level, $g_{i}$ the (arbitrary) number of levels
in the $i$th cell, and, in a particular situation, $n_{i}$ is the number of
particles in the $i$th cell. Moreover, $a=-1$ for the Bose--Einstein case,
$+1$ for the Fermi--Dirac case, and $0$ for the (classical) Maxwell--Boltzmann
case. Even if nothing is said explicitly about the set $I$ of the indices $i,$
from several examples in the literature, $I$ is a countable set; the example%

\[
\varepsilon(n_{x},n_{y},n_{z})=\frac{h^{2}}{8mL^{2}}(n_{x}^{2}+n_{y}^{2}%
+n_{z}^{2});\quad n_{x},n_{y},n_{z}=1,2,3,\ldots
\]
is considered in \cite[p.\ 10]{PatBea:11}.

Relation (\ref{PB}) suggests the consideration of the following functions
defined on $\mathbb{R}$ with values in $\overline{\mathbb{R}}$, called,
respectively, Bose--Einstein, Fermi--Dirac and Maxwell--Boltzmann, entropies:%
\begin{equation}
E_{BE}(u):=\left\{
\begin{array}
[c]{ll}%
u\ln u-(1+u)\ln(1+u) & \text{if }u\in\mathbb{R}_{+},\\
\infty & \text{if }u\in\mathbb{R}_{-}^{\ast},
\end{array}
\right.  \label{ebe}%
\end{equation}%
\begin{equation}
E_{FD}(u):=\left\{
\begin{array}
[c]{ll}%
u\ln u+(1-u)\ln(1-u) & \text{if }u\in\lbrack0,1],\\
\infty & \text{if }u\in\mathbb{R}\setminus\lbrack0,1],
\end{array}
\right.  \label{efd}%
\end{equation}%
\begin{equation}
E_{MB}(u):=\left\{
\begin{array}
[c]{ll}%
u(\ln u-1) & \text{if }u\in\mathbb{R}_{+},\\
\infty & \text{if }u\in\mathbb{R}_{-}^{\ast},
\end{array}
\right.  \label{emb}%
\end{equation}
where $0\ln0:=0$ and $\mathbb{R}_{+}:=[0,\infty\lbrack$, $\mathbb{R}_{+}%
^{\ast}:={}]0,\infty\lbrack$, $\mathbb{R}_{-}:=-\mathbb{R}_{+}$,
$\mathbb{R}_{-}^{\ast}:=-\mathbb{R}_{+}^{\ast}$. We have that
\[
E_{BE}^{\prime}(u)=\ln\frac{u}{1+u}~\forall u\in\mathbb{R}_{+}^{\ast},\quad
E_{FD}^{\prime}(u)=\ln\frac{u}{1-u}~\forall u\in{}]0,1[,\quad E_{MB}^{\prime
}(u)=\ln u~\forall u\in\mathbb{R}_{+}^{\ast}.
\]
Observe that $E_{BE}$, $E_{MB}$, $E_{FD}$ are convex (even strictly convex on
their domains), derivable on the interiors of their domains with increasing
derivatives, and $E_{BE}\leq E_{MB}\leq E_{FD}$ on $\mathbb{R}$. The (convex)
conjugates of these functions are%
\[
E_{MB}^{\ast}(t)=e^{t}~\forall t\in\mathbb{R},\quad E_{FD}^{\ast}%
(t)=\ln(1+e^{t})~\forall t\in\mathbb{R},\quad E_{BE}^{\ast}(t)=\left\{
\begin{array}
[c]{ll}%
-\ln(1-e^{t}) & \text{if }t\in\mathbb{R}_{-}^{\ast},\\
\infty & \text{if }t\in\mathbb{R}_{+}.
\end{array}
\right.
\]
Moreover, for $W\in\{E_{BE},E_{MB},E_{FD}\}$ we have that $\partial
W(u)=\{W^{\prime}(u)\}$ for $u\in\operatorname*{int}(\operatorname*{dom}W)$
and $\partial W(u)=\emptyset$ elsewhere; furthermore,
\begin{equation}
(W^{\ast})^{\prime}(t)=\frac{e^{t}}{1+a_{W}e^{t}}\quad\forall t\in
\operatorname*{dom}W^{\ast}, \label{r-mbf4}%
\end{equation}
where (as above)
\begin{equation}
a_{W}:=\left\{
\begin{array}
[c]{ll}%
-1 & \text{if }W=E_{BE},\\
0 & \text{if }W=E_{MB},\\
1 & \text{if }W=E_{FD}.
\end{array}
\right.  \label{r-aw}%
\end{equation}

The maximization of (\ref{PB}) subject to the constraints $\sum_{i}n_{i}=N$
and $\sum_{i}n_{i}\varepsilon_{i}=E$ is equivalent to the minimization problem

\medskip

minimize $\sum_{i}g_{i}W(\frac{n_{i}}{g_{i}})$ \ \ s.t. \ $\sum_{i}n_{i}=N,$
\ $\sum_{i}n_{i}\varepsilon_{i}=E,$

\medskip

\noindent where $W$ is one of the functions $E_{BE},\ E_{FD},\ E_{MB}$ defined
in (\ref{ebe}), (\ref{efd}), (\ref{emb}), and $g_{i}\geq1$.

In many books treating this subject (see \cite[pp.\ 119, 120]{LanLif:80},
\cite[pp.\ 15, 16]{Gue:07}, \cite[p.\ 144]{PatBea:11}, \cite[p.\ 39]{Bas:14})
the above problem is solved using the Lagrange multipliers method in a formal way.

Our aim is to treat rigorously the minimization of Maxwell--Boltzmann,
Bose--Einstein and Fermi--Dirac entropies with the constraints $\sum_{i\in
I}u_{i}=u$, $\sum_{i\in I}\sigma_{i}u_{i}=v$ in the case in which $I$ is a
countable set. Unfortunately, we succeed to do a complete study only for the
Maxwell--Boltzmann entropy. For a short description of the results see Conclusions.

Our approach is based on the results of X. Y. Zheng \cite{Zhe:98} on the
subdifferential of a countable sum of convex functions and on our recent
results in \cite{ValZal:15}\footnote{See the preprint arXiv:1506.01216v1.} for
the conjugate of such a function.

We shall use standard notations and results from convex analysis (see
e.g.\ \cite{Roc:70}, \cite{Zal:02}).

\section{Properties of the marginal functions associated to the entropy
minimization problems of Statistical Mechanics}

Throughout the paper we consider the sequences $(p_{n})_{n\geq1}\subset
\lbrack1,\infty\lbrack$ and $(\sigma_{n})_{n\geq1}\subset\mathbb{R}$, and set
\begin{equation}
S(u,v):=S_{(\sigma_{n})}(u,v):=\bigg\{(u_{n})_{n\geq1}\subset\mathbb{R}%
_{+}\mid u=\sum\nolimits_{n\geq1}u_{n},\ v=\sum\nolimits_{n\geq1}\sigma
_{n}u_{n}\bigg\} \label{r-su}%
\end{equation}
for each $(u,v)\in\mathbb{R}^{2}$. It is clear that $S(u,v)=S(tu,tv)$ for all
$(u,v)\in\mathbb{R}^{2}$ and $t\in\mathbb{R}_{+}^{\ast}$, $S(u,v)=\emptyset$
if either $u<0$ or $u=0\neq v$, and $S(0,0)=\{(0)_{n\geq1}\}$. We also set%
\begin{gather}
\rho_{n}:=\sum\nolimits_{k=1}^{n}p_{k}\label{r-ron}\\
\eta_{n}^{1}:=\min\left\{  \sigma_{k}\mid k\in\overline{1,n}\right\}
,\quad\eta_{n}^{2}:=\max\left\{  \sigma_{k}\mid k\in\overline{1,n}\right\}
,\label{r-etn}\\
\eta_{1}:=\inf\left\{  \sigma_{k}\mid n\geq1\right\}  \in\lbrack-\infty
,\infty\lbrack,\quad\eta_{2}:=\sup\left\{  \sigma_{k}\mid n\geq1\right\}
\in{}]\!-\infty,\infty]; \label{r-et12}%
\end{gather}
of course, $\lim_{n\rightarrow\infty}\rho_{n}=\infty$ (because $p_{k}\geq1$
for $n\geq1$).

The entropy minimization problem (EMP for short) of Statistical Mechanics and
Statistical Physics associated to $W\in\{E_{BE}$, $E_{MB}$, $E_{FD}\}$ and
$(u,v)\in\mathbb{R}^{2}$ is

\medskip

$(EMP)_{u,v}$ minimize $\sum_{n\geq1}p_{n}W(\frac{u_{n}}{p_{n}})$ s.t.
$(u_{n})_{n\geq1}\in S(u,v),$

\medskip

\noindent where $\sum_{n\geq1}\beta_{n}:=\lim_{n\rightarrow\infty}\sum
_{k=1}^{n}\beta_{k}$ when this limit exists in $\overline{\mathbb{R}}$ and
$\sum_{n\geq1}\beta_{n}:=\infty$ otherwise. With the preceding convention, it
is easy to see that $\alpha_{n}\leq\beta_{n}$ for $n\geq1$ imply that
$\sum_{n\geq1}\alpha_{n}\leq\sum_{n\geq1}\beta_{n}.$

\begin{remark}
\label{rem1}Note that for $(u_{n})_{n\geq1}\in S(u,v)$ one has that
$\lim_{n\rightarrow\infty}\sum_{k=1}^{n}p_{k}W(\frac{u_{k}}{p_{k}})$ exists
in: $[-\infty,0]$ when $W=E_{BE}$, in $[-\infty,0]\cup\{\infty\}$ when
$W=E_{FD}$, and in $[-\infty,\infty\lbrack$ when $W=E_{MB}.$
\end{remark}

The value (marginal) function associated to problems $(EMP)_{u,v}$ is
\begin{equation}
H_{W}:\mathbb{R}^{2}\rightarrow\overline{\mathbb{R}},\quad H_{W}%
(u,v):=\inf\bigg\{\sum\nolimits_{n\geq1}p_{n}W\bigg(\frac{u_{n}}{p_{n}%
}\bigg)\mid(u_{n})_{n\geq1}\in S(u,v)\bigg\}, \label{r-hu}%
\end{equation}
with the usual convention $\inf\emptyset:=\infty$. We shall write
$H_{(\sigma_{n}),W}^{(p_{n})}$ instead of $H_{W}$ when we want to emphasize
the sequences $(p_{n})_{n\geq1}$ and $(\sigma_{n})_{n\geq1}$; moreover, we
shall write simply $H_{BE}$, $H_{MB}$, $H_{FD}$ when $W$ is $E_{BE}$, $E_{MB}%
$, or $E_{FD}$, respectively. Therefore,
\[
\operatorname*{dom}H_{W}=\operatorname*{dom}H_{(\sigma_{n}),W}^{(p_{n}%
)}\subset\operatorname*{dom}S=\operatorname*{dom}S_{(\sigma_{n})}:=\left\{
(u,v)\in\mathbb{R}^{2}\mid S_{(\sigma_{n})}(u,v)\neq\emptyset\right\}
;\label{r-dhw}%
\]
hence $H_{W}(u,v)=\infty$ if either $u<0$ or $u=0\neq v$, and $H_{W}(0,0)=0$.
Taking into account that $E_{BE}\leq E_{MB}\leq E_{FD}$, and using Remark
\ref{rem1}, we get
\begin{gather}
H_{BE}\leq H_{MB}\leq H_{FD},\label{r-hes}\\
\operatorname*{dom}H_{FD}\subset\operatorname*{dom}H_{MB}=\operatorname*{dom}%
H_{BE}=\operatorname*{dom}S. \label{r-dhs}%
\end{gather}

The results in the next two lemmas are surely known. For their proofs one uses
the Lagrange multipliers method.

\begin{lemma}
\label{lem2}Let $n\geq2$ be fixed. Then for $W\in\{E_{BE}$, $E_{MB}$,
$E_{FD}\}$ we have
\begin{equation}
\inf\bigg\{\sum_{k=1}^{n}p_{k}W\left(  \frac{u_{k}}{p_{k}}\right)  \mid
(u_{k})_{k\in\overline{1,n}}\subset\mathbb{R}_{+},\ \sum_{k=1}^{n}%
u_{k}=u\bigg\}=\rho_{n}\cdot W\left(  u/\rho_{n}\right)  \quad\forall
u\in\mathbb{R}_{+}, \label{r-wfin}%
\end{equation}
the infimum being attained for $u_{k}:=up_{k}/\rho_{n}$ $(k\in\overline{1,n}%
)$, where $\rho_{n}$ is defined in (\ref{r-ron}).
\end{lemma}

Proof. Consider
\begin{equation}
\widetilde{W}_{n}:\mathbb{R}^{n}\rightarrow\overline{\mathbb{R}}%
,\quad\widetilde{W}_{n}(u_{1},\ldots,u_{n}):=\sum\nolimits_{k=1}^{n}%
p_{k}W\left(  \frac{u_{k}}{p_{k}}\right)  . \label{r-wt}%
\end{equation}
Then $\operatorname*{dom}\widetilde{W}_{n}=\mathbb{R}_{+}^{n}$ for
$W\in\{E_{MB},E_{BE}\}$ and $\operatorname*{dom}\widetilde{W}_{n}=%
{\textstyle\prod_{k=1}^{n}}
[0,p_{k}]$ for $W=E_{FD}$. Of course, $\widetilde{W}_{n}$ is convex, lower
semi\-continuous (lsc for short), continuous on $\operatorname*{int}%
(\operatorname*{dom}\widetilde{W}_{n})$, and strictly convex on
$\operatorname*{dom}\widetilde{W}_{n}$. Let $S_{n}^{\prime}(u):=\big\{(u_{1}%
,\ldots u_{n})\in\mathbb{R}_{+}^{n}\mid\sum_{k=1}^{n}u_{k}=u\big\}$. Since
$S_{n}^{\prime}(0)=\{(0)_{k\in\overline{1,n}}\}$, the conclusion is obvious
for $u=0.$

Consider first $W\in\{E_{MB},E_{BE}\}$, and take $u\in\mathbb{R}_{+}^{\ast}$.
Then $u\rho_{n}^{-1}(p_{1},\ldots,p_{n})\in S_{n}^{\prime}(u)\cap
\operatorname*{int}(\operatorname*{dom}\widetilde{W}_{n})$. Since
$S_{n}^{\prime}(u)$ is a compact set and $\widetilde{W}_{n}$ is lsc, there
exists a unique $(\overline{u}_{1},\ldots,\overline{u}_{n})\in S_{n}^{\prime
}(u)$ minimizing $\widetilde{W}_{n}$ on $S_{n}^{\prime}(u)$. Using (for
example) \cite[Th.\ 2.9.6]{Zal:02}, there exists $\alpha\in\mathbb{R}$ such
that $\alpha(1,\ldots,1)\in\partial\widetilde{W}_{n}(\overline{u}_{1}%
,\ldots,\overline{u}_{n})=\partial W\big(\frac{\overline{u}_{1}}{p_{1}%
}\big)\times\ldots\times\partial W\big(\frac{\overline{u}_{n}}{p_{n}}\big)$.
Since $\partial W(0)=\emptyset$, it follows that $\overline{u}_{k}/p_{k}>0$
for $k\in\overline{1,n}$. Hence $\partial W\big(\frac{\overline{u}_{k}}{p_{k}%
}\big)=\big\{W^{\prime}\big(\frac{\overline{u}_{k}}{p_{k}}\big)\big\}$ for
$k\in\overline{1,n}$, whence $\frac{\overline{u}_{k}}{p_{k}}=:\eta$. Thus,
$u=\sum_{k=1}^{n}\overline{u}_{k}=\eta\rho_{n}$, that is $\eta=u/\rho_{n}$,
and so $\overline{u}_{k}=up_{k}/\rho_{n}$ for $k\in\overline{1,n}$. It follows
that the infimum in (\ref{r-wfin}) is $\rho_{n}\cdot W\left(  u/\rho
_{n}\right)  $.

Consider now $W=E_{FD}$. For $u=\rho_{n}$ we have that $S_{n}^{\prime}%
(\rho_{n})=\{p\}$, where $p:=(p_{1},\ldots,p_{n})$, and $S_{n}^{\prime
}(u)=\emptyset$ for $u>\rho_{n}$; hence, the conclusion is trivial for
$u\geq\rho_{n}.$

Let $u\in{}]0,\rho_{n}[$. Then $u\rho_{n}^{-1}p\in S_{n}^{\prime}%
(u)\cap\operatorname*{int}(\operatorname*{dom}\widetilde{W}_{n})$. The rest of
the proof is the same as that of the preceding case. The proof is complete.
$\quad\square$

\medskip

For $(u,v)\in\mathbb{R}^{2}$ and $n\geq1$ let us set
\[
S_{n}^{\prime\prime}(u,v):=\bigg\{(u_{1},\ldots u_{n})\in\mathbb{R}_{+}%
^{n}\mid\sum\nolimits_{k=1}^{n}u_{k}=u,\ \sum\nolimits_{k=1}^{n}u_{k}%
\sigma_{k}=v\bigg\},
\]
and
\[
\operatorname*{dom}S_{n}^{\prime\prime}:=\left\{  (u,v)\in\mathbb{R}^{2}\mid
S_{n}^{\prime\prime}(u,v)\neq\emptyset\right\}  =T_{n}(\mathbb{R}_{+}^{n}),
\]
where
\begin{equation}
T_{n}:\mathbb{R}^{n}\rightarrow\mathbb{R}^{2},\quad T_{n}(u_{1},\ldots
,u_{n}):=\sum\nolimits_{k=1}^{n}u_{k}(1,\sigma_{k}). \label{r-tn}%
\end{equation}
It follows that
\begin{align}
\operatorname*{dom}S_{n}^{\prime\prime}  &  =\sum\nolimits_{k=1}^{n}%
\mathbb{R}_{+}\cdot(1,\sigma_{k})=\mathbb{R}_{+}(1,\eta_{n}^{1})+\mathbb{R}%
_{+}(1,\eta_{n}^{2})\nonumber\\
&  =\left\{  (u,v)\in\mathbb{R}_{+}\times\mathbb{R}\mid\eta_{n}^{1}u\leq
v\leq\eta_{n}^{2}u\right\}  , \label{r-dhsn}%
\end{align}
where $\eta_{n}^{1}$, $\eta_{n}^{2}$ are defined in (\ref{r-etn}). Hence
$\operatorname*{ri}(\operatorname*{dom}S_{n}^{\prime\prime})=\mathbb{R}%
_{+}^{\ast}(1,\eta_{n}^{1})$ when $\eta_{n}^{1}=\eta_{n}^{2}$; if $\eta
_{n}^{1}<\eta_{n}^{2}$ then $T_{n}$ is surjective, and so%
\begin{align}
\operatorname*{int}(\operatorname*{dom}S_{n}^{\prime\prime})  &
=\operatorname*{int}\left(  T_{n}(\mathbb{R}_{+}^{n})\right)  =T_{n}\left(
\operatorname*{int}\mathbb{R}_{+}^{n}\right)  =\mathbb{R}_{+}^{\ast}%
(1,\eta_{n}^{1})+\mathbb{R}_{+}^{\ast}(1,\eta_{n}^{2})\nonumber\\
&  =\left\{  (u,v)\in\mathbb{R}_{+}^{\ast}\times\mathbb{R}\mid\eta_{n}%
^{1}u<v<\eta_{n}^{2}u\right\}  \text{.} \label{r-intsns}%
\end{align}
Observe that $S_{n}^{\prime\prime}(u,v)=\mathbb{R}_{+}^{n}\cap T_{n}%
^{-1}\left(  \{(u,v)\}\right)  $ and $S_{n}^{\prime\prime}(u,v)$ is convex and
compact for each $(u,v)\in\operatorname*{dom}S_{n}^{\prime\prime}.$

In the next result we characterize the solutions of the minimization problem

\medskip

$(EMP)_{u,v}^{n}$ minimize $\widetilde{W}_{n}(u_{1},\ldots,u_{n})$ s.t.
$(u_{k})_{k\in\overline{1,n}}\in S_{n}^{\prime\prime}(u,v),$

\medskip

\noindent where $\widetilde{W}_{n}$ is defined in (\ref{r-wt}).

\begin{lemma}
\label{lem3a}Let $n\geq2$ be fixed, and $W\in\{E_{BE}$, $E_{MB}\}$. Assume
that $\eta_{n}^{1}<\eta_{n}^{2}$ [see (\ref{r-etn})] and take $(u,v)\in
\mathbb{R}_{+}\times\mathbb{R}$ such that $\eta_{n}^{1}u\leq v\leq\eta_{n}%
^{2}v$. Then $(EMP)_{u,v}^{n}$ has a unique solution $(\overline{u}_{1}%
,\ldots,\overline{u}_{n})$. Moreover, the following assertions are true:

\emph{(i)} If $u=0$ then $\overline{u}_{k}=0$ for every $k\in\overline{1,n}.$

\emph{(ii)} If $\eta_{n}^{1}u<v<\eta_{n}^{2}u$ then $\overline{u}_{k}>0$ for
every $k\in\overline{1,n}$. Moreover, there exist (and they are unique)
$\alpha,\beta\in\mathbb{R}$ such that $W^{\prime}(\overline{u}_{k}%
/p_{k})=\alpha+\beta\sigma_{k}$ for every $k\in\overline{1,n}.$

\emph{(iii)} If $u\in\mathbb{R}_{+}^{\ast}$ and $v=\eta_{n}^{i}u$ for some
$i\in\{1,2\}$, then $\Sigma_{i}:=\{k\in\overline{1,n}\mid\sigma_{k}=\eta
_{n}^{i}\}\neq\emptyset$ and $\overline{u}_{k}=up_{k}/\sum_{l\in\Sigma_{i}%
}p_{l}$ for $k\in\Sigma_{i}$, $\overline{u}_{k}=0$ for $k\in\overline
{1,n}\setminus\Sigma_{i}.$
\end{lemma}

Proof. Since $S_{n}^{\prime\prime}(u,v)$ $(\subset\operatorname*{dom}%
\widetilde{W}_{n}=\mathbb{R}_{+}^{n})$ is a nonempty [see (\ref{r-dhsn})]
compact set and $\widetilde{W}_{n}$ is lsc and strictly convex on
$\operatorname*{dom}\widetilde{W}_{n}$, $(EMP)_{u,v}^{n}$ has a unique
solution $(\overline{u}_{1},\ldots,\overline{u}_{n})$.

(i) The assertion is obvious.

(ii) By (\ref{r-intsns}) we have that $(u,v)\in\operatorname*{int}%
(\operatorname*{dom}S_{n}^{\prime\prime})=T_{n}\big(\operatorname*{int}%
(\operatorname*{dom}\widetilde{W}_{n})\big)$. Using again \cite[Th.\ 2.9.6]%
{Zal:02}, there exist $\alpha,\beta\in\mathbb{R}$ such that
\[
\alpha(1,\ldots,1)+\beta(\sigma_{1},\ldots,\sigma_{n})\in\partial\widetilde
{W}_{n}(\overline{u}_{1},\ldots,\overline{u}_{n})=\partial W\big(\overline
{u}_{1}/p_{1}\big)\times\ldots\times\partial W\big(\overline{u}_{n}%
/p_{n}\big).
\]
It follows that $\overline{u}_{k}/p_{k}>0$, and so $\alpha+\beta\sigma
_{k}=W^{\prime}\big(\frac{\overline{u}_{k}}{p_{k}}\big)$ for $k\in
\overline{1,n}$. Since $\eta_{n}^{1}\neq\eta_{n}^{2}$, $\alpha$ and $\beta$
are unique.

(iii) Consider the case $v=\eta_{n}^{1}u$ with $u\in\mathbb{R}_{+}^{\ast}$
(the case $i=2$ being similar). Take $(u_{1},\ldots,u_{n})\in S_{n}%
^{\prime\prime}(u,v)$. Then $\sum_{k=1}^{n}u_{k}(\sigma_{k}-\eta_{n}^{1})=0$,
whence $u_{k}(\sigma_{k}-\eta_{n}^{1})=0$ for every $k\in\overline{1,n}$. It
follows that $u_{k}=0$ for $k\in\overline{1,n}\setminus\Sigma_{1}$. Therefore,
problem $(EMP)_{u,v}^{n}$ is equivalent to minimizing $\sum_{k\in\Sigma_{1}%
}p_{k}W\big(\frac{u_{k}}{p_{k}}\big)$ with the constraint $\sum_{k\in
\Sigma_{1}}u_{k}=u$. Using Lemma \ref{lem2}, the unique solution of this
problem is $(\overline{u}_{k})_{k\in\Sigma_{1}}$ with $\overline{u}_{k}%
=up_{k}/\sum_{l\in\Sigma_{1}}p_{l}$. $\quad\square$

\medskip

The argument for the proof of the next result is very similar to that in the
proof of the preceding one, so we omit it.

\begin{lemma}
\label{lem3b}Let $n\geq2$ be fixed and $W=E_{FD}$. Assume that $(u,v)\in
\mathbb{R}_{+}\times\mathbb{R}$ is such that $S_{n}^{\prime\prime}%
(u,v)\cap\operatorname*{dom}\widetilde{W}_{n}\neq\emptyset$. Then
$(EMP)_{u,v}^{n}$ has a unique solution $(\overline{u}_{1},\ldots,\overline
{u}_{n})$. Moreover, if $S_{n}^{\prime\prime}(u,v)\cap\operatorname*{int}%
(\operatorname*{dom}\widetilde{W}_{n})\neq\emptyset$, then there exist (and
they are unique) $\alpha,\beta\in\mathbb{R}$ such that $W^{\prime}%
(\overline{u}_{k}/p_{k})=\alpha+\beta\sigma_{k}$ for every $k\in\overline
{1,n}$; in particular, $(\overline{u}_{1},\ldots,\overline{u}_{n}%
)\in\operatorname*{int}(\operatorname*{dom}\widetilde{W}_{n}).$
\end{lemma}

Note that $\alpha$ and $\beta$ from Lemma \ref{lem3a} in case $W=E_{MB}$ can
be obtained (quite easily) using Lemma \ref{lem1} (below); indeed,
$\beta=(\varphi_{n})^{-1}(v/u)$ and $\alpha=\ln\left(  u/\sum_{k=1}^{n}%
p_{k}e^{\sigma_{k}\beta}\right)  .$

\begin{lemma}
\label{lem1}Let $n\geq2$ and suppose that $\eta_{n}^{1}<\eta_{n}^{2}$ (see
(\ref{r-et12})). Consider the function
\begin{equation}
\varphi_{n}:\mathbb{R}\rightarrow\mathbb{R},\quad\varphi_{n}(t):=\frac
{\sum_{k=1}^{n}p_{k}\sigma_{k}e^{\sigma_{k}t}}{\sum_{k=1}^{n}p_{k}%
e^{\sigma_{k}t}}.\label{r-fin}%
\end{equation}
Then $\varphi_{n}$ is increasing and $\lim_{t\rightarrow-\infty}\varphi
_{n}(t)=\eta_{n}^{1}$, $\lim_{t\rightarrow\infty}\varphi_{n}(t)=\eta_{n}^{2}.$
Therefore, $\varphi_{n}(\mathbb{R})={}]\eta_{n}^{1},\eta_{n}^{2}[.$
\end{lemma}

Proof. We have that
\[
\varphi_{n}^{\prime}(t):=\frac{\sum_{k=1}^{n}p_{k}\sigma_{k}^{2}e^{\sigma
_{k}t}\cdot\sum_{k=1}^{n}p_{k}e^{\sigma_{k}t}-\left(  \sum_{k=1}^{n}%
p_{k}\sigma_{k}e^{\sigma_{k}t}\right)  ^{2}}{\left(  \sum_{k=1}^{n}%
p_{k}e^{\sigma_{k}t}\right)  ^{2}}\quad\forall t\in\mathbb{R}.
\]
By Cauchy--Bunyakovsky inequality we have that
\[
\bigg(\sum_{k=1}^{n}\left[  \sigma_{k}\left(  p_{k}e^{\sigma_{k}t}\right)
^{\frac{1}{2}}\right]  \cdot\left(  p_{k}e^{\sigma_{k}t}\right)  ^{\frac{1}%
{2}}\bigg)^{2}<\sum_{k=1}^{n}p_{k}\sigma_{k}^{2}e^{\sigma_{k}t}\cdot\sum
_{k=1}^{n}p_{k}e^{\sigma_{k}t}\quad\forall t\in\mathbb{R}%
\]
(the inequality being strict because $\eta_{n}^{1}<\eta_{n}^{2}$), and so
$\varphi_{n}^{\prime}(t)>0$ for every $t\in\mathbb{R}.$

Set $\Sigma_{i}:=\{k\in\overline{1,n}\mid\sigma_{k}=\eta_{n}^{i}\}$ for
$i\in\{1,2\}$. Since $\lim_{t\rightarrow-\infty}e^{\sigma t}=0$ for $\sigma
\in\mathbb{R}_{+}^{\ast}$, we obtain that
\[
\lim_{t\rightarrow-\infty}\varphi_{n}(t)=\lim_{t\rightarrow-\infty}\frac
{\sum_{k=1}^{n}p_{k}\sigma_{k}e^{(\sigma_{k}-\eta_{n}^{1})t}}{\sum_{k=1}%
^{n}p_{k}e^{(\sigma_{k}-\eta_{n}^{1})t}}=\frac{\sum_{k\in\Sigma_{1}}%
p_{k}\sigma_{k}}{\sum_{k\in\Sigma_{1}}p_{k}}=\eta_{n}^{1}.
\]
Similarly, $\lim_{t\rightarrow\infty}\varphi_{n}(t)=\eta_{n}^{2}$. Because
$\varphi_{n}$ is increasing and continuous we obtain that $\varphi
_{n}(\mathbb{R})={}]\eta_{n}^{1},\eta_{n}^{2}[{}$. $\quad\square$

\medskip

In the next result we establish the convexity of $H_{W}$ and estimate its
domain for $W\in\{E_{MB},E_{BE},E_{FD}\}.$

\begin{proposition}
\label{p1}Let $\eta_{1}$ and $\eta_{2}$ be defined in (\ref{r-et12}). Set
$S:=S_{(\sigma_{n})}$ and $H_{W}:=H_{(\sigma_{n}),W}^{(p_{n})}$ for
$W\in\left\{  E_{BE},E_{MB},E_{FD}\right\}  $. The following assertions hold:

\emph{(i)} The marginal function $H_{W}$ is convex; moreover, (\ref{r-hes})
and (\ref{r-dhs}) hold.

\emph{(ii)} Assume that $\eta_{1}=\eta_{2}$. Then $\operatorname*{dom}%
H_{W}=\operatorname*{dom}S=\mathbb{R}_{+}\cdot(1,\sigma_{1})$; in particular,
$\operatorname*{ri}(\operatorname*{dom}H_{W})=\mathbb{R}_{+}^{\ast}%
(1,\sigma_{1})\neq\emptyset=\operatorname*{int}(\operatorname*{dom}H_{W}).$

\emph{(iii) }Assume that $\eta_{1}<\eta_{2}$ and take $\overline{n}\geq2$ such
that $\{\sigma_{k}\mid k\in\overline{1,\overline{n}}\}$ is not a singleton.
Then for $W\in\{E_{BE},E_{MB}\}$ one has
\begin{gather}
C:=\bigcup_{n\geq1}\sum_{k=1}^{n}\mathbb{R}_{+}\cdot(1,\sigma_{k}%
)\subset\operatorname*{dom}H_{W}=\operatorname*{dom}S\subset\operatorname*{cl}%
C,\label{r-dommb}\\
\operatorname*{int}(\operatorname*{dom}H_{W})=\operatorname*{int}%
(\operatorname*{dom}S)=\operatorname*{int}C=\bigcup_{n\geq\overline{n}}%
\sum_{k=1}^{n}\mathbb{R}_{+}^{\ast}\cdot(1,\sigma_{k})=\mathbb{R}_{+}^{\ast
}\cdot\left(  \{1\}\times{}]\eta_{1},\eta_{2}[{}\right)  . \label{r-dommb2}%
\end{gather}
Moreover,%
\begin{gather}
A:=\bigcup_{n\geq1}\sum_{k=1}^{n}[0,p_{k}]\cdot(1,\sigma_{k})\subset
\operatorname*{dom}H_{FD}\subset\operatorname*{cl}A,\label{r-domfd}\\
\operatorname*{int}(\operatorname*{dom}H_{FD})=\operatorname*{int}%
A=\bigcup_{n\geq\overline{n}}\sum_{k=1}^{n}{}]0,p_{k}[{}\cdot(1,\sigma_{k}).
\label{r-domfd2}%
\end{gather}

\end{proposition}

Proof. (i) Let $(u,v)$, $(u^{\prime},v^{\prime})\in\operatorname*{dom}H_{W}$
and $\lambda\in{}]0,1[$. Take $\mu,\mu^{\prime}\in\mathbb{R}$ with
$H_{W}(u,v)<\mu$, $H_{W}(u^{\prime},v^{\prime})<\mu^{\prime}$; there exist
$(u_{n})_{n\geq1}\in S(u,v)$, $(u_{n}^{\prime})_{n\geq1}\in S(u^{\prime
},v^{\prime})$ such that $\sum_{n\geq1}p_{n}W\big(\frac{u_{n}}{p_{n}}%
\big)<\mu$, $\sum_{n\geq1}p_{n}W\big(\frac{u_{n}^{\prime}}{p_{n}}%
\big)<\mu^{\prime}$. Clearly, $\left(  \lambda u_{n}+(1-\lambda)u_{n}^{\prime
}\right)  _{n\geq1}\in S\big(\lambda(u,v)+(1-\lambda)(u^{\prime},v^{\prime
})\big)$. Since $p_{n}W\big(\frac{\lambda u_{n}+(1-\lambda)u_{n}^{\prime}%
}{p_{n}}\big)\leq\lambda p_{n}W\big(\frac{u_{n}}{p_{n}}\big)+(1-\lambda
)p_{n}W\big(\frac{u_{n}^{\prime}}{p_{n}}\big)$, summing up term by term for
$n\geq1$, we get
\[
H_{W}\big(\lambda(u,v)+(1-\lambda)(u^{\prime},v^{\prime})\big)\leq\lambda
\sum_{n\geq1}p_{n}W\big(\frac{u_{n}}{p_{n}}\big)+(1-\lambda)\sum_{n\geq1}%
p_{n}W\big(\frac{u_{n}^{\prime}}{p_{n}}\big)<\lambda\mu+(1-\lambda)\mu
^{\prime}.
\]
Letting $\mu\rightarrow H_{W}(u,v)$ and $\mu^{\prime}\rightarrow
H_{W}(u^{\prime},v^{\prime})$ we get $H_{W}\big(\lambda(u,v)+(1-\lambda
)(u^{\prime},v^{\prime})\big)\leq\lambda H_{W}(u,v)+(1-\lambda)H_{W}%
(u^{\prime},v^{\prime})$. Hence $H_{W}$ is convex.

(ii) Assume that $\eta_{1}=\eta_{2}$; hence $\sigma_{n}=\sigma_{1}$ for
$n\geq1$. Then $\operatorname*{dom}S=\mathbb{R}_{+}\cdot(1,\sigma_{1})$.

Taking into account (\ref{r-dhs}), it is sufficient to show that
$\operatorname*{dom}S\subset\operatorname*{dom}H_{FD}$. Take $u\in
\mathbb{R}_{+}$. If $u<p_{1}$, take $u_{1}:=u$ and $u_{k}:=0$ for $n\geq2$. If
$u\geq p_{1}$, there exists $n\geq1$ such that $\rho_{n}:=\sum_{k=1}^{n}%
p_{k}\leq u<\rho_{n+1}$. Take $u_{k}=p_{k}$ for $k\in\overline{1,n}$,
$u_{n+1}:=u-\rho_{n}<p_{n+1}$, $u_{k}:=0$ for $k\geq n+1$. In both cases we
have that $u=\sum_{k\geq1}u_{k}$ and $\sum_{k\geq1}p_{k}W(u_{k}/p_{k})\leq0$,
and so $(u,u\sigma_{1})\in\operatorname*{dom}H_{W}$. Hence
$\operatorname*{dom}S=\mathbb{R}_{+}\cdot(1,\sigma_{1})\subset
\operatorname*{dom}H_{FD}.$

From the expression of $\operatorname*{dom}H_{W}$, the last assertion is obvious.

(iii) Assume that $\eta_{1}<\eta_{2}$. The first three inclusions in
(\ref{r-dommb}) are obvious. For the last one, take $(u,v)\in
\operatorname*{dom}S$; then there exists $(u_{n})_{n\geq1}\in S(u,v)$. Since
$C\ni\sum_{k=1}^{n}u_{k}(1,\sigma_{k})\rightarrow(u,v)$, we have that
$(u,v)\in\operatorname*{cl}C.$

Set $C_{n}:=\sum_{k=1}^{n}\mathbb{R}_{+}\cdot(1,\sigma_{k})$. Clearly
$C_{n}\subset C_{n+1}$ and $C=\mathfrak{\cup}_{n\geq1}C_{n}$; hence $C$ is
convex. The first two equalities in (\ref{r-dommb2}) follow from
(\ref{r-dommb}) because $C$ is convex. Take $n\geq\overline{n}$. Since the
linear operator $T_{n}:\mathbb{R}^{n}\rightarrow\mathbb{R}^{2}$ defined in
(\ref{r-tn}) is surjective and $C_{n}=T_{n}\left(  \mathbb{R}_{+}^{n}\right)
$, we have that
\[
\operatorname*{int}C_{n}=T_{n}\big(\operatorname*{int}\mathbb{R}_{+}%
^{n}\big)=T_{n}\big({{\prod}_{k=1}^{n}} \mathbb{R}_{+}^{\ast}\big)={\sum
}_{k=1}^{n}\mathbb{R}_{+}^{\ast}\cdot(1,\sigma_{k}).
\]
Since $(C_{n})_{n\geq\overline{n}}$ is an increasing sequence of convex sets
with nonempty interior and $C=\cup_{n\geq\overline{n}}C_{n}$, we obtain that
$\operatorname*{int}C=\cup_{n\geq\overline{n}}\operatorname*{int}C_{n}$. Hence
the third equality in (\ref{r-dommb2}) holds.

Take now $(u,v)\in\sum_{k=1}^{n}\mathbb{R}_{+}^{\ast}\cdot(1,\sigma_{k})$ for
some $n\geq\overline{n}$. Then $(u,v)=\sum_{k=1}^{n}u_{k}(1,\sigma_{k})$ with
$(u_{k})_{k\in\overline{1,n}}\subset\mathbb{R}_{+}^{\ast}$. It follows that
$(u,v)=\alpha\cdot(1,w)$, where $\alpha:=\sum_{i=1}^{n}u_{i}\in\mathbb{R}%
_{+}^{\ast}$ and $w:=\sum_{k=1}^{n}\frac{u_{k}}{\alpha}\sigma_{k}$. Because
$\{\sigma_{k}\mid k\in\overline{1,n}\}$ is not a singleton, $w\in{}]\eta
_{1},\eta_{2}[$, and so $(u,v)\in B:=\mathbb{R}_{+}^{\ast}\cdot\left(
\{1\}\times{}]\eta_{1},\eta_{2}[{}\right)  $; clearly, $B$ is open.
Conversely, take $(u,v)\in B$, that is $(u,v)=\alpha\cdot(1,w)$ with
$\alpha\in\mathbb{R}_{+}^{\ast}$ and $w\in{}]\eta_{1},\eta_{2}[$. Then there
exists $n_{1},n_{2}\geq2$ such that $\sigma_{n_{1}}<w<\sigma_{n_{2}}$; hence
$w=\lambda\sigma_{n_{1}}+(1-\lambda)\sigma_{n_{2}}$ for some $\lambda\in
{}]0,1[$. Consider $n=\max\{\overline{n},n_{1},n_{2}\}$. It follows that
$(u,v)\in C_{n}\subset C$. Therefore, $B\subset C,$ whence
$B=\operatorname*{int}B\subset\operatorname*{int}C$. Hence the last equality
in (\ref{r-dommb2}) holds, too.

The first inclusion in (\ref{r-domfd}) is obvious. Take $(u,v)\in
\operatorname*{dom}H_{FD}$; then there exists $(u_{n})_{n\geq1}\in S(u,v)$
such that $\sum_{n\geq1}p_{n}H_{FD}\big(\frac{u_{n}}{p_{n}}\big)<\infty$. It
follows that $u_{n}\in\lbrack0,p_{n}]$ for $n\geq1$. Since $A\ni\sum_{k=1}%
^{n}u_{k}(1,\sigma_{k})\rightarrow(u,v)$, we have that $(u,v)\in
\operatorname*{cl}A$. Hence (\ref{r-domfd}) holds.

Set $A_{n}:=\sum_{k=1}^{n}[0,p_{k}]\cdot(1,\sigma_{k})$. Clearly $A_{n}\subset
A_{n+1}$ and $A=\mathfrak{\cup}_{n\geq1}A_{n}$, and so $A$ is convex. The
first equality in (\ref{r-domfd2}) follows from (\ref{r-domfd}) because $A$ is
convex. Take $n\geq\overline{n}$. Since the linear operator $T_{n}$ is
surjective and $A_{n}=T_{n}\left(
{\textstyle\prod_{k=1}^{n}}
[0,p_{k}]\right)  $, we have that
\[
\operatorname*{int}A_{n}=T_{n}\big(\operatorname*{int}%
{\textstyle\prod_{k=1}^{n}}
[0,p_{k}]\big)=T_{n}\big(%
{\textstyle\prod_{k=1}^{n}}
]0,p_{k}[\big)={\sum}_{k=1}^{n}{}]0,p_{k}[{}\cdot(1,\sigma_{k}).
\]
Since $(A_{n})_{n\geq\overline{n}}$ is an increasing sequence of convex sets
with nonempty interior and $A=\cup_{n\geq\overline{n}}A_{n}$, we obtain that
$\operatorname*{int}A=\cup_{n\geq\overline{n}}\operatorname*{int}A_{n}.$
Therefore, the last equality in (\ref{r-domfd2}) holds, too. The proof is
complete. $\quad\square$

\begin{proposition}
\label{p2}Consider $W\in\{E_{BE},E_{MB},E_{FD}\}.$

\emph{(i)} Assume that $\eta_{1}=\eta_{2}$. Then $H_{W}(u,v)=-\infty$ for all
$(u,v)\in\operatorname*{ri}(\operatorname*{dom}H_{W})=\mathbb{R}_{+}^{\ast
}\cdot(1,\sigma_{1})$.

\emph{(ii)} Assume that the series $\sum_{n\geq1}p_{n}e^{\sigma_{n}x}$ is
divergent for every $x\in\mathbb{R}$ and $\eta_{1}<\eta_{2}$. Then
\begin{equation}
H_{W}(u,v)=-\infty~~\forall(u,v)\in\operatorname*{int}(\operatorname*{dom}%
H_{W}). \label{r-hesi}%
\end{equation}

\end{proposition}

Proof. (i) Clearly, $\sigma_{n}=\sigma_{1}$ for $n\geq1$. Take $u>0$. Then
there exists $\overline{n}\geq1$ such that $u<\rho_{n}=\sum_{k=1}^{n}p_{k}$
for $n\geq\overline{n}$. Having in view Lemma \ref{lem2}, consider
$u_{k}:=up_{k}/\rho_{n}$ for $k\in\overline{1,n}$ and $u_{k}:=0$ for $k\geq
n+1$. Then $u=\sum_{k\geq1}u_{k}$, and so%
\[
H_{FD}(u,u\sigma_{1})\leq{\sum}_{k=1}^{n}p_{k}E_{FD}\left(  u/\rho_{n}\right)
=u\ln u-u\ln\rho_{n}-u\left(  1-\frac{u}{\rho_{n}}\right)  \frac
{\ln\big(1-\frac{u}{\rho_{n}}\big)}{-\frac{u}{\rho_{n}}}%
\]
for all $n\geq\overline{n}$. Since $\rho_{n}\rightarrow\infty$, we obtain that
$H_{FD}(u,u\sigma_{1})=-\infty$. Using (\ref{r-hes}), we have that
$H_{W}(u,v)=-\infty$ for $(u,v)\in\operatorname*{ri}(\operatorname*{dom}%
H_{W})=\mathbb{R}_{+}^{\ast}\cdot(1,\sigma_{1})$ for $W\in\left\{
E_{BE},E_{MB},E_{FD}\right\}  .$

(ii) Take first $W=E_{MB}$. Because $S_{(\sigma_{\phi(n)})}(u,v)=\{(u_{\phi
(n)})\mid(u_{n})\in S_{(\sigma_{n})}(u,v)\}$ and $H_{(\sigma_{n}),W}^{(p_{n}%
)}(u,v)=H_{(\sigma_{\phi(n)}),W}^{(p_{\phi(n)})}(u,v)$ for every bijection
$\phi:\mathbb{N}^{\ast}\rightarrow\mathbb{N}^{\ast}$ with $\phi(n)=n$ for
large $n$, we may (and do) assume that $\sigma_{1}<\sigma_{2}$. Even more,
because for $a\in\mathbb{R}$ and $\sigma_{n}^{\prime}:=\sigma_{n}-a$
$(n\geq1)$ we have that $S_{(\sigma_{n})}(u,v)=S_{(\sigma_{n}^{\prime}%
)}(u,v-au)$ and, consequently, $H_{(\sigma_{n}),W}^{(p_{n})}(u,v)=H_{(\sigma
_{n}^{\prime}),W}^{(p_{n})}(u,v-au)$, we may (and do) assume that $0\in
{}]\sigma_{1},\sigma_{2}[$. For $n\geq2$ consider the function $\varphi_{n}$
defined in (\ref{r-fin}). By Lemma \ref{lem1} there exists (a unique)
$y_{n}\in\mathbb{R}$ such that $\varphi_{n}(y_{n})=0$. We claim that
$\lim_{n\rightarrow\infty}\sum_{k=1}^{n}p_{k}e^{\sigma_{k}y_{n}}=\infty$. In
the contrary case there exist an increasing sequence $(n_{m})_{m\geq1}%
\subset\mathbb{N}^{\ast}\setminus\{1,2\}$ and $M\in\mathbb{R}_{+}^{\ast}$ such
that $\sum_{k=1}^{n_{m}}p_{k}e^{\sigma_{k}y_{n_{m}}}\leq M$ for every $m\geq
1$. In particular, $p_{1}e^{\sigma_{1}y_{n_{m}}}\leq M$ (whence $y_{n_{m}}%
\geq(\ln M-\ln p_{1})/\sigma_{1}$) and $e^{\sigma_{2}y_{n_{m}}}\leq M$ (whence
$y_{n_{m}}\leq(\ln M-\ln p_{2})/\sigma_{2}$) for $m\geq1$; hence $(y_{n_{m}%
})_{m\geq1}$ is bounded. Passing if necessary to a subsequence, we may (and
do) assume that $y_{n_{m}}\rightarrow y\in\mathbb{R}$. For $q\geq2$ there
exists $m_{q}\geq1$ such that $n_{m}\geq q$ for every $m\geq m_{q}$. Hence
$\sum_{k=1}^{q}p_{k}e^{\sigma_{k}y_{n_{m}}}\leq M$ for every $m\geq m_{q}$.
Letting $(m_{q}\leq)$ $m\rightarrow\infty$ we obtain that $\sum_{k=1}^{q}%
p_{k}e^{\sigma_{k}y}\leq M$ for every $q\geq2$, and so we get the
contradiction that the series $\sum_{k\geq1}p_{k}e^{\sigma_{k}y}$ is
convergent. Therefore, our claim is true.

Set $x_{n}:=-\ln\left(  \sum_{k=1}^{n}p_{k}e^{\sigma_{k}y_{n}}\right)
\rightarrow-\infty$ for $n\rightarrow\infty$. Set $u_{k}:=p_{k}e^{x_{n}%
+\sigma_{k}y_{n}}>0$ for $k\in\overline{1,n}$ and $u_{k}:=0$ for $k\geq n+1.$
Then
\[
\sum\nolimits_{k\geq1}u_{k}=\sum\nolimits_{k=1}^{n}p_{k}e^{x_{n}+\sigma
_{k}y_{n}}=1,\quad\sum\nolimits_{k\geq1}u_{k}\sigma_{k}=\sum\nolimits_{k=1}%
^{n}p_{k}\sigma_{k}e^{x_{n}+\sigma_{k}y_{n}}=0,
\]
and so $\left(  u_{k}\right)  _{k\geq1}\in S(1,0)$. Hence $u_{k}=p_{k}%
e^{x_{n}+\sigma_{k}y_{n}}\leq1\leq p_{k}$, and so $e^{x_{n}+\sigma_{k}y_{n}%
}\leq1$, for every $k\in\overline{1,n}$. It follows that
\begin{align*}
H_{FD}(1,0)  &  \leq\sum_{k\geq1}p_{k}E_{FD}\big(\frac{u_{k}}{p_{k}}%
\big)=\sum_{k=1}^{n}u_{k}\ln e^{x_{n}+\sigma_{k}y_{n}}+\sum_{k=1}^{n}%
(p_{k}-u_{k})\ln(1-e^{x_{n}+\sigma_{k}y_{n}})\\
&  \leq\sum_{k=1}^{n}u_{k}(x_{n}+\sigma_{k}y_{n})=x_{n}\quad\forall n\geq2,
\end{align*}
and so $H_{FD}(1,0)=-\infty$. Using (\ref{r-hes}) we obtain that
(\ref{r-hesi}) holds. The proof is complete. $\quad\square$

\medskip

The previous result shows the lack of interest of the EMP when the sequence
$(\sigma_{n})_{n\geq1}$ is constant. Also, it gives a hint on the importance
of knowing the properties of the function
\begin{equation}
f:\mathbb{R}\rightarrow\overline{\mathbb{R}},\quad f(x)={\sum}_{n\geq1}%
p_{n}e^{\sigma_{n}x}. \label{r-f}%
\end{equation}

The next result, with $p_{n}=1$ for $n\geq1$, is practically \cite[Prop.
12]{ValZal:15}; the adaptation of its proof for the present case is easy.

\begin{proposition}
\label{prop1}Let $f_{n}(x):=p_{n}e^{\sigma_{n}x}$ for $n\geq1$, $x\in
\mathbb{R}$, and set $f=\sum_{n\geq1}f_{n}.$

\emph{(i)} If $\overline{x}\in\operatorname*{dom}f$ then $\sigma_{n}%
\overline{x}\rightarrow-\infty$, and so either $\overline{x}>0$ and
$\sigma_{n}\rightarrow-\infty$, or $\overline{x}<0$ and $\sigma_{n}%
\rightarrow\infty.$

Furthermore, assume that $(A_{\sigma f})$ holds, where

\medskip

$(A_{\sigma f})$ $~~(\sigma_{n})_{n\geq1}\subset\mathbb{R}_{+}^{\ast},$
$\sigma_{n}\rightarrow\infty$, and $\operatorname*{dom}f\neq\emptyset.$

\medskip

\emph{(ii)} Then there exists $\alpha\in\mathbb{R}_{+}$ such that $I:={}%
]{}-\infty,-\alpha\lbrack{}\subset\operatorname*{dom}f\subset\mathbb{R}%
_{-}^{\ast}\cap\operatorname*{cl}I$, $f$ is strictly convex and increasing on
$\operatorname*{dom}f$, and $\lim_{x\rightarrow-\infty}f(x)=0=\inf f.$
Moreover,
\[
f^{\prime}(x)=\sum\nolimits_{n\geq1}f_{n}^{\prime}(x)=\sum\nolimits_{n\geq
1}p_{n}\sigma_{n}e^{\sigma_{n}x}\quad\forall x\in\operatorname*{int}%
(\operatorname*{dom}f)=I, \label{r-8}%
\]
$f^{\prime}$ is increasing and continuous on $I$, $\lim_{x\rightarrow-\infty
}f^{\prime}(x)=0$, and
\[
\lim_{x\uparrow-\alpha}f^{\prime}(x)=\sum\nolimits_{n\geq1}p_{n}\sigma
_{n}e^{-\sigma_{n}\alpha}=:\gamma\in{}]0,\infty]. \label{r-7}%
\]
In particular, $\partial f(\operatorname*{int}(\operatorname*{dom}%
f))=f^{\prime}(I)={}]0,\gamma\lbrack.$

\emph{(iii)} Let $\alpha$, $I$, $\gamma$ be as in \emph{(ii)}. Assume that
$\alpha\in\mathbb{R}_{+}^{\ast}$. Then either \emph{(a) }$\operatorname*{dom}%
f=I$ and $\gamma=\infty$, or \emph{(b)} $\operatorname*{dom}%
f=\operatorname*{cl}I$ and $\gamma=\infty$, in which case $f_{-}^{\prime
}(-\alpha)=\gamma$, $\partial f(-\alpha)=\emptyset$ and the series
$\sum_{n\geq1}f_{n}^{\prime}(-\alpha)$ is divergent, or \emph{(c)}
$\operatorname*{dom}f=\operatorname*{cl}I$ and $\gamma<\infty$, in which case
$f_{-}^{\prime}(-\alpha)=\gamma$ and%
\[
\sum\nolimits_{n\geq1}f_{n}^{\prime}(-\alpha)=\gamma\in\lbrack\gamma
,\infty\lbrack{}=\partial f(-\alpha).
\]

\end{proposition}

Let us consider the following functions for $W\in\{E_{MB},E_{FD},E_{BE}\}:$
\begin{gather*}
h_{n}^{W}:\mathbb{R}^{2}\rightarrow\overline{\mathbb{R}},\quad h_{n}%
^{W}(x,y):=p_{n}W^{\ast}(x+\sigma_{n}y)>0\quad(n\geq1,\ x,y\in\mathbb{R}),\\
h_{W}:=h_{W,(\sigma_{n})}^{(p_{n})}:\mathbb{R}^{2}\rightarrow\overline
{\mathbb{R}},\quad h_{W}:=\sum\nolimits_{n\geq1}h_{n}^{W};
\end{gather*}
we write simply $h_{MB}$, $h_{FD}$, $h_{BE}$ instead of $h_{E_{MB}}$,
$h_{E_{FD}}$, $h_{E_{BE}}$, respectively. Because $h_{n}^{W}=(p_{n}W^{\ast
})\circ A_{n}$, where $A_{n}:\mathbb{R}^{2}\rightarrow\mathbb{R}$ is defined
by $A_{n}(x,y):=x+\sigma_{n}y$ [and so $A_{n}^{\ast}w=w(1,\sigma_{n})$], we
have that
\begin{align}
\left(  h_{n}^{W}\right)  ^{\ast}(u,v) &  =\min\left\{  (p_{n}W^{\ast})^{\ast
}(w)\mid A_{n}^{\ast}w=(u,v)\right\}  =\min\big\{p_{n}W\big(\frac{w}{p_{n}%
}\big)\mid A_{n}^{\ast}w=(u,v)\big\}\nonumber\\
&  =\left\{
\begin{array}
[c]{ll}%
p_{n}W(\frac{u}{p_{n}}) & \text{if }u\geq0\text{ and }v=\sigma_{n}u,\\
\infty & \text{otherwise,}%
\end{array}
\right.  \label{r-hwns}%
\end{align}
and so $\left(  h_{n}^{W}\right)  ^{\ast}$ is strictly convex on its domain.

The expression of $\left(  h_{n}^{W}\right)  ^{\ast}$ (above) in connection
with \cite[Prop.\ 15(i)]{ValZal:15} shows the interest of studying the
properties of the functions $h_{W}.$

\section{Properties of the functions $h_{W}$}

Because $p_{n}\geq1$ for $n\geq1$, we have that
\begin{align}
(x,y)\in\operatorname*{dom}h_{W}  &  \Rightarrow p_{n}W^{\ast}(x+\sigma
_{n}y)\rightarrow0\Rightarrow W^{\ast}(x+\sigma_{n}y)\rightarrow
0\Leftrightarrow\sigma_{n}y\rightarrow-\infty\nonumber\\
&  \Leftrightarrow\lbrack y>0\text{ and }\sigma_{n}\rightarrow-\infty]\text{
or }[y<0\text{ and }\sigma_{n}\rightarrow\infty]. \label{r-mbf1}%
\end{align}
Of course, $h_{n}^{W}(x,y)>0$ for all $(x,y)\in\mathbb{R}^{2}$, $n\geq1$ and
$W\in\{E_{MB},E_{FD},E_{BE}\}$; because for $\sigma_{n}y\rightarrow-\infty$ we
have
\[
\lim_{n\rightarrow\infty}\frac{h_{n}^{E_{FD}}(x,y)}{h_{n}^{E_{MB}}(x,y)}%
=\lim_{n\rightarrow\infty}\frac{h_{n}^{E_{BE}}(x,y)}{h_{n}^{E_{MB}}(x,y)}=1,
\]
we obtain that
\[
\operatorname*{dom}h_{FD}=\operatorname*{dom}h_{MB},\quad\operatorname*{dom}%
h_{BE}=\operatorname*{dom}h_{MB}\cap\left\{  (x,y)\in\mathbb{R}^{2}\mid
x+\sigma_{n}y<0~\forall n\geq1\right\}  .
\]

Since $h_{MB}(x,y)=\sum_{n\geq1}p_{n}e^{x+\sigma_{n}y}=e^{x}f(y)$, where $f$
is defined by (\ref{r-f}), clearly $\operatorname*{dom}h_{MB}=\mathbb{R}%
\times\operatorname*{dom}f$. It follows that
\[
\operatorname*{dom}h_{FD}\neq\emptyset\Leftrightarrow\operatorname*{dom}%
h_{MB}\neq\emptyset\Leftrightarrow\operatorname*{dom}h_{BE}\neq\emptyset
\Leftrightarrow\operatorname*{dom}f\neq\emptyset.
\]

Because it is natural to consider the case in which $\operatorname*{dom}h_{W}$
is nonempty, in the sequel, we assume that $(A_{\sigma f})$ holds. (The
general case can be reduced to this one replacing $(\sigma_{n})_{n\geq1}$ by
$(-\sigma_{n})_{n\geq1}$ if $\sigma_{n}\rightarrow-\infty$, then replacing
$(\sigma_{n})_{n\geq1}$ by $(\sigma_{n}-a)_{n\geq1}$ with $a<\min_{n\geq
1}\sigma_{n}$.)

\begin{proposition}
\label{p-mbf1}Assume that $(A_{\sigma f})$ holds, and take $\alpha
\in\mathbb{R}_{+}$ such that $I:={}]\!-\infty,-\alpha\lbrack{}\subset
\operatorname*{dom}f\subset\operatorname*{cl}I$, where $f$ is defined in
(\ref{r-f}). Let $W\in\{E_{MB},E_{FD},E_{BE}\}.$

\emph{(i)} Then $h_{W}$ is convex, lower semi\-continuous, positive, and
\[
\operatorname*{dom}h_{FD}=\operatorname*{dom}h_{MB}=\mathbb{R}\times
\operatorname*{dom}f,\quad\operatorname*{dom}h_{BE}=\left\{  (x,y)\in
\mathbb{R}\times\operatorname*{dom}f\mid x+\theta_{1}y<0\right\}  ,
\]
where $\theta_{1}:=\min\{\sigma_{n}\mid n\geq1\}$.

\emph{(ii)} $h_{W}$ is differentiable at any $(x,y)\in\operatorname*{int}%
(\operatorname*{dom}h_{W})$ and
\begin{equation}
\nabla h_{W}(x,y)=\sum\nolimits_{n\geq1}\nabla h_{n}^{W}(x,y)=\sum
\nolimits_{n\geq1}p_{n}\frac{e^{x+\sigma_{n}y}}{1+a_{W}e^{x+\sigma_{n}y}}%
\cdot(1,\sigma_{n}), \label{r-mbf3}%
\end{equation}
$a_{W}$ being defined in (\ref{r-aw}). Moreover, assume that $(x,-\alpha
)\in\operatorname*{dom}h_{W}$; in particular, $-\alpha\in\operatorname*{dom}%
f\subset\mathbb{R}_{-}^{\ast}$. Then
\begin{equation}
\partial h_{W}(x,-\alpha)\neq\emptyset\iff\sum_{n\geq1}\nabla h_{n}%
^{W}(x,-\alpha)\text{ converges}\iff\gamma:=\sum_{n\geq1}p_{n}\sigma
_{n}e^{x-\sigma_{n}\alpha}\in\mathbb{R}; \label{r-mbf8}%
\end{equation}
if $(\overline{u},\overline{v}):=\sum\nolimits_{n\geq1}\nabla h_{n}%
^{W}(x,-\alpha)$ exists in $\mathbb{R}^{2}$, then
\begin{equation}
\partial h_{W}(x,-\alpha)=\{\overline{u}\}\times\lbrack\overline{v}%
,\infty\lbrack{}=\left\{  (\overline{u},\overline{v})\right\}  +\{0\}\times
\mathbb{R}_{+}. \label{r-mbf7}%
\end{equation}

\end{proposition}

Proof. The existence of $\alpha\in\mathbb{R}_{+}$ such that $I:={}%
]\!-\infty,-\alpha\lbrack{}\subset\operatorname*{dom}f\subset
\operatorname*{cl}I$ is ensured by Proposition \ref{prop1}.

(i) Since $h_{MB}(x,y)=e^{x}f(y)$ for $(x,y)\in\mathbb{R}^{2}$, we have that
$\operatorname*{dom}h_{MB}=\mathbb{R}\times\operatorname*{dom}f$. Taking into
account (\ref{r-mbf1}) and the fact that $\lim_{t\rightarrow-\infty}%
e^{-t}W(t)=1$ for $W\in\{E_{MB},E_{FD},E_{BE}\}$, we obtain that
\begin{equation}
\operatorname*{dom}h_{W}=\left(  \mathbb{R}\times\operatorname*{dom}f\right)
\cap%
{\textstyle\bigcap\nolimits_{n\geq1}}
\operatorname*{dom}h_{n}^{W}\subset\left(  \mathbb{R}\times\mathbb{R}%
_{-}^{\ast}\right)  \cap%
{\textstyle\bigcap\nolimits_{n\geq1}}
\operatorname*{dom}h_{n}^{W}. \label{r-mbf2}%
\end{equation}
Since $\operatorname*{dom}h_{n}^{W}=\mathbb{R}^{2}$ for $n\geq1$ and
$W\in\{E_{MB},E_{FD}\}$, we get $\operatorname*{dom}h_{W}=\mathbb{R}%
\times\operatorname*{dom}f$ for $W\in\{E_{MB},E_{FD}\}$. Let $W=E_{BE}$; then
$\operatorname*{dom}h_{n}^{W}=\{(x,y)\in\mathbb{R}^{2}\mid x+\sigma_{n}y<0\}.$
For $x\in\mathbb{R}$ and $y\in\operatorname*{dom}f$ (hence $y<0$), $(x,y)\in%
{\textstyle\bigcap\nolimits_{n\geq1}}
\operatorname*{dom}h_{n}^{W}$ if and only if $x+\theta_{1}y<0$. From
(\ref{r-mbf2}) we get the given expression of $\operatorname*{dom}h_{BE}$.

Since $h_{n}^{W}$ is convex, continuous and positive, we obtain that $h_{W}$
is convex, lower semi\-continuous and positive.

(ii) Using \cite[Cor.\ 11]{ValZal:15} (and (\ref{r-mbf4})), we obtain that
$h_{W}$ is differentiable on $\operatorname*{int}(\operatorname*{dom}h_{W})$
and (\ref{r-mbf3}) holds.

Assume that $(x,-\alpha)\in\operatorname*{dom}h_{W}$. Hence $-\alpha
\in\operatorname*{dom}f\subset\mathbb{R}_{-}^{\ast};$ moreover, $x\in
\mathbb{R}$ for $W\in\{E_{MB},E_{FD}\}$, and $x<\theta_{1}\alpha$ $(\leq
\sigma_{n}\alpha$ for $n\geq1)$ for $W=E_{BE}$.

Because $\lim_{n\rightarrow\infty}\left(  1+a_{W}e^{x+\sigma_{n}y}\right)
=1$, the last equivalence in (\ref{r-mbf8}) holds.

In order to prove the first equivalence in (\ref{r-mbf8}), suppose first that
$(u,v)\in\partial h_{W}(x,-\alpha)$. Then, for $(x^{\prime},y)\in
\operatorname*{dom}h_{W},$
\begin{equation}
u\cdot(x^{\prime}-x)+v\cdot(y+\alpha)\leq h_{W}(x^{\prime},y)-h_{W}%
(x,-\alpha).\label{r-mbf6}%
\end{equation}
Taking $y:=-\alpha$ we obtain that $u\in\partial h_{W}(\cdot,-\alpha)(x).$
Since $h_{W}(\cdot,-\alpha)=\sum_{n\geq1}h_{n}^{W}(\cdot,-\alpha)$ and
$x\in\operatorname*{int}(\operatorname*{dom}h_{W}(\cdot,-\alpha))$, using
again \cite[Cor.\ 11]{ValZal:15} we obtain that $h_{W}(\cdot,-\alpha)$ is
derivable at $x$ and
\[
u=\left(  h_{W}(\cdot,-\alpha)\right)  ^{\prime}(x)=\sum\nolimits_{n\geq
1}\left(  h_{n}^{W}(\cdot,-\alpha)\right)  ^{\prime}(x).\label{r-mbf5}%
\]
Take now $x^{\prime}=x$ and $y<-\alpha$ in (\ref{r-mbf6}). Dividing by
$y+\alpha$ $(<0)$, and taking into account that $W^{\ast}$ is increasing on
its domain, we get
\[
v\geq\sum_{n\geq1}p_{n}\frac{W^{\ast}(x+\sigma_{n}y)-W^{\ast}(x-\sigma
_{n}\alpha)}{y+\alpha}\geq\sum_{k=1}^{n}p_{k}\frac{W^{\ast}(x+\sigma
_{k}y)-W^{\ast}(x-\sigma_{k}\alpha)}{y+\alpha}\quad\forall n\geq1.
\]
Taking the limit for $y\uparrow-\alpha$ in the second inequality, we get
$v\geq\sum_{k=1}^{n}p_{k}\sigma_{k}(W^{\ast})^{\prime}(x-\sigma_{k}\alpha).$
Since $(W^{\ast})^{\prime}(x-\sigma_{k}\alpha)>0$ for every $k\geq1$, we
obtain that the series $\sum_{n\geq1}p_{n}\sigma_{n}(W^{\ast})^{\prime
}(x-\sigma_{n}\alpha)=\sum\nolimits_{n\geq1}\left(  h_{n}^{W}(x,\cdot)\right)
^{\prime}(-\alpha)$ is convergent and $v\geq\sum\nolimits_{n\geq1}\left(
h_{n}^{W}(x,\cdot)\right)  ^{\prime}(-\alpha)$. Hence the series
$\sum\nolimits_{n\geq1}\nabla h_{n}^{W}(x,-\alpha)$ is convergent and for its
sum $(\overline{u},\overline{v})$ we have that $u=\overline{u}$ and
$v\geq\overline{v}.$

Conversely, assume that the series $\sum\nolimits_{n\geq1}\nabla h_{n}%
^{W}(x,-\alpha)$ is convergent with sum $(\overline{u},\overline{v})$; take
$v\geq\overline{v}$ and $(x^{\prime},y)\in\operatorname*{dom}h_{W}$. Using
\cite[Prop.\ 15(iii)]{ValZal:15} we obtain that $(\overline{u},\overline
{v})\in\partial h_{W}(x,-\alpha)$. Hence (\ref{r-mbf6}) holds for $(u,v)$
replaced by $(\overline{u},\overline{v})$. Since $v\geq\overline{v}$ and
$y\leq-\alpha$, we have that $v\cdot(y+\alpha)\leq\overline{v}\cdot(y+\alpha
)$, and so (\ref{r-mbf6}) also holds for $u$ replaced by $\overline{u}$. It
follows that $(\overline{u},v)\in\partial h_{W}(x,-\alpha)$. Therefore,
(\ref{r-mbf7}) holds. $\quad\square$

\begin{theorem}
\label{t-mbf1}Let $W\in\{E_{MB},E_{FD},E_{BE}\}$ and $a_{W}$ be defined in
(\ref{r-aw}). Then for every $(x,y)\in\cap_{n\geq1}\operatorname*{dom}%
h_{n}^{W}$ such that the series $\sum_{n\geq1}p_{n}\frac{e^{x+\sigma_{n}y}%
}{1+a_{W}e^{x+\sigma_{n}y}}\cdot(1,\sigma_{n})$ is convergent [this is the
case, for example, when $(x,y)\in\operatorname*{int}(\operatorname*{dom}%
h_{W})$] with sum $(u,v)\in\mathbb{R}^{2}$, the problem $(EMP)_{u,v}$ has the
unique optimal solution $\left(  p_{n}\frac{e^{x+\sigma_{n}y}}{1+a_{W}%
e^{x+\sigma_{n}y}}\right)  _{n\geq1}$. Moreover, the value of the problem
$(EMP)_{u,v}$ is $h_{W}^{\ast}(u,v)$, that is $H_{W}(u,v)=h_{W}^{\ast}(u,v).$
\end{theorem}

Proof. Taking into account that $h_{n}^{W}$ is a proper convex function for
$n\geq1$ with
\[
\partial h_{n}^{W}(x,y)=\{\nabla h_{n}^{W}(x,y)\}=\{p_{n}(W^{\ast})^{\prime
}(e^{x+\sigma_{n}y})\cdot(1,\sigma_{n})\}=\left\{  p_{n}\frac{e^{x+\sigma
_{n}y}}{1+a_{W}e^{x+\sigma_{n}y}}\cdot(1,\sigma_{n})\right\}
\]
for $(x,y)\in\operatorname*{dom}h_{n}^{W}$, and $h_{W}=\sum_{n\geq1}h_{n}^{W}%
$, as well as the expression $\left(  h_{n}^{W}\right)  ^{\ast}$ given in
(\ref{r-hwns}), we get the conclusion using \cite[Prop.\ 15(iii)]{ValZal:15}.
The fact that the series $\sum_{n\geq1}p_{n}\frac{e^{x+\sigma_{n}y}}%
{1+a_{W}e^{x+\sigma_{n}y}}\cdot(1,\sigma_{n})$ is convergent for
$(x,y)\in\operatorname*{int}(\operatorname*{dom}h_{W})$ is ensured by
\cite[Cor.\ 11(i)]{ValZal:15}.$\quad\square$

\medskip

The result in Theorem \ref{t-mbf1} is obtained generally using the Lagrange
multipliers method in a formal way.

A complete solution to EMP for the Maxwell--Boltzmann entropy is provided in
the next section.

\section{Complete solution to EMP in the case of the Maxwell--Boltzmann
entropy}

In this section $W=E_{MB}$; to simplify the writing, we set $h_{n}%
:=h_{n}^{E_{MB}}$, $h:=h_{MB}$, $H:=H_{MB}$; we mention also the sequences
$(\sigma_{n})$ and $(p_{n})$ if necessary. We assume that $(A_{\sigma f})$
holds if not stated explicitly otherwise. Moreover, we use $I$, $\alpha$,
$\gamma$ as in Proposition \ref{prop1}.

From (\ref{r-hwns}) we have that%

\[
h_{n}^{\ast}(u,v)=\left\{
\begin{array}
[c]{ll}%
u(\ln\frac{u}{p_{n}}-1) & \text{if }u\geq0\text{ and }v=\sigma_{n}u,\\
\infty & \text{otherwise,}%
\end{array}
\right.
\]
and so $h_{n}^{\ast}$ is strictly convex on its domain.

Let us compute $h^{\ast}$. From the definition of the conjugate, for
$(u,v)\in\mathbb{R}^{2}$ we have that
\begin{align*}
h^{\ast}(u,v)  &  =\sup_{(x,y)\in\mathbb{R}^{2}}\left[  ux+vy-e^{x}%
f(y)\right]  =\sup_{y\in\operatorname*{dom}f}\left[  vy+\sup_{x\in\mathbb{R}%
}[ux-e^{x}f(y)]\right] \\
&  =\sup_{y\in\operatorname*{dom}f}\bigg[vy+f(y)\sup_{x\in\mathbb{R}%
}\bigg(x\frac{u}{f(y)}-e^{x}\bigg)\bigg]=\sup_{y\in\operatorname*{dom}%
f}\bigg[vy+f(y)\exp^{\ast}\bigg(\frac{u}{f(y)}\bigg)\bigg].
\end{align*}
It follows that $h^{\ast}(u,v)=\infty$ for $u\in\mathbb{R}_{-}^{\ast}$ and
$h^{\ast}(0,v)=\iota_{\operatorname*{dom}f}^{\ast}(v)$; hence $h^{\ast
}(0,v)=\infty$ for $v\in\mathbb{R}_{-}^{\ast}$ and $h^{\ast}(0,v)=-\alpha v$
for $v\in\mathbb{R}_{+}.$

Fix $(u,v)\in\mathbb{R}_{+}^{\ast}\times\mathbb{R}$. From the above expression
of $h^{\ast}(u,v)$ we get%
\begin{align*}
h^{\ast}(u,v)  &  =\sup_{y\in I}\left[  vy+u\ln u-u\ln\left[  f(y)\right]
-u\right]  =u(\ln u-1)+u\sup_{y\in I}\left[  \frac{v}{u}y-\ln\left[
f(y)\right]  \right] \\
&  =u(\ln u-1)+u\cdot(\ln f)^{\ast}(v/u).
\end{align*}

Below we show that $\ln f:=\ln\circ f:\mathbb{R}\rightarrow\overline
{\mathbb{R}}$ is convex; we also calculate its conjugate. For these consider
\begin{equation}
\varphi:I\rightarrow\mathbb{R}_{+}^{\ast},\quad\varphi(y):=f^{\prime}(y)/f(y).
\label{r-21}%
\end{equation}
Observe, using Schwarz' inequality in $\ell_{2}$, that
\[
\left[  f^{\prime}(y)\right]  ^{2}=\bigg[\sum_{n\geq1}\sigma_{n}%
(p_{n}e^{\sigma_{n}y})^{\frac{1}{2}}\cdot(p_{n}e^{\sigma_{n}y})^{\frac{1}{2}%
}\bigg]^{2}<\sum_{n\geq1}\sigma_{n}^{2}p_{n}e^{\sigma_{n}y}\sum_{n\geq1}%
p_{n}e^{\sigma_{n}y}=f(y)\cdot f^{\prime\prime}(y)\quad\forall y\in I
\]
(the inequality being strict because $\sigma_{n}\rightarrow\infty$). Hence
$\varphi^{\prime}(y)=\big(f(y)\cdot f^{\prime\prime}(y)-[f^{\prime}%
(y)]^{2}\big)/[f(y)]^{2}>0$ for $y\in I$. Therefore, $\varphi$ is increasing,
and so
\[
0<\theta_{1}:=\min_{n\geq1}\sigma_{n}=\lim_{y\rightarrow-\infty}%
\varphi(y)<\lim_{y\uparrow-\alpha}\varphi(y)=:\theta_{2}\leq\infty.
\]
Restricting the co-domain of $\varphi$ to $]\theta_{1},\theta_{2}[$ we get an
increasing bijective function denoted also by $\varphi$. Since $(\ln
f)^{\prime}=f^{\prime}/f=\varphi$ on $I$, $\ln f$ is (strictly) convex on its
domain. Observe that $\theta_{2}<\infty$ is equivalent to $-\alpha
\in\operatorname*{dom}f$ and $\gamma:=f_{-}^{\prime}(-\alpha)<\infty$, in
which case $\theta_{2}=f_{-}^{\prime}(-\alpha)/f(-\alpha)$. Indeed, assume
that $\theta_{2}<\infty$ and fix $y_{0}\in I$. Then for $y_{0}<y<-\alpha$ we
have that $\ln\frac{f(y)}{f(y_{0})}=\int_{y_{0}}^{y}\varphi(t)dt\leq\theta
_{2}(y-y_{0})$, whence $f(y)\leq f(y_{0})e^{\theta_{2}(y-y_{0})}$, and so
$f(-\alpha)\leq f(y_{0})e^{-\theta_{2}(\alpha+y_{0})}<\infty$; then, because
$f^{\prime}(y)=f(y)\cdot\varphi(y)$ for $y\in I$, we get $\gamma=\theta
_{2}\cdot f(-\alpha)$. The converse implication is obvious. Because
$\operatorname*{int}(\operatorname*{dom}(\ln f))=I$, it follows that
\[
(\ln f)^{\ast}\left(  w\right)  =\sup_{y\in I}\big(wy-\ln\left[  f(y)\right]
\big)=\left\{
\begin{array}
[c]{ll}%
\infty & \text{if }w<\theta_{1},\\
w\varphi^{-1}(w)-\ln\left[  f(\varphi^{-1}(w))\right]  & \text{if }\theta
_{1}<w<\theta_{2,}\\
-\alpha w-\ln\left[  f(-\alpha)\right]  & \text{if }\theta_{2}\leq w,
\end{array}
\right.
\]
where the last line has to be taken into consideration only if $\theta
_{2}<\infty$. Let us set
\[
\Sigma:=\{k\in\mathbb{N}^{\ast}\mid\sigma_{k}=\theta_{1}\};
\]
of course, $\Sigma$ is finite and nonempty, and so we may (and do) suppose
that $\Sigma=\overline{1,q}$ for some $q\in\mathbb{N}^{\ast}$ and
$\sigma_{q+1}\leq\sigma_{n}$ for $n\geq q+1$. Because $(\ln f)^{\ast}$ is
convex and lsc, we have that
\[
(\ln f)^{\ast}\left(  \theta_{1}\right)  =\lim_{w\downarrow\theta_{1}}(\ln
f)^{\ast}\left(  w\right)  =\lim_{w\downarrow\theta_{1}}\big[w\varphi
^{-1}(w)-\ln[f(\varphi^{-1}(w))]\big]=\lim_{y\rightarrow-\infty}\psi(y),
\]
where $\psi(y)=y\varphi(y)-\ln\left[  f(y)\right]  $ for $y\in I$. But
\begin{align*}
\psi(y)  &  =y\frac{\sigma_{1}\sum_{n=1}^{q}p_{n}+\sum_{n\geq q+1}p_{n}%
\sigma_{n}e^{(\sigma_{n}-\sigma_{1})y}}{\sum_{n=1}^{q}p_{n}+\sum_{n\geq
q+1}p_{n}e^{(\sigma_{n}-\sigma_{1})y}}-\sigma_{1}y-\ln\bigg(\sum_{n=1}%
^{q}p_{n}+\sum_{n\geq q+1}p_{n}e^{(\sigma_{n}-\sigma_{1})y}\bigg)\\
&  =ye^{(\sigma_{q+1}-\sigma_{1})y}\cdot\frac{\sum_{n\geq q+1}p_{n}(\sigma
_{n}-\sigma_{1})e^{(\sigma_{n}-\sigma_{q+1})y}}{\sum_{n=1}^{q}p_{n}%
+\sum_{n\geq q+1}p_{n}e^{(\sigma_{n}-\sigma_{1})y}}-\ln\bigg(\sum_{n=1}%
^{q}p_{n}+\sum_{n\geq q+1}p_{n}e^{(\sigma_{n}-\sigma_{1})y}\bigg).
\end{align*}
It follows that $(\ln f)^{\ast}\left(  \theta_{1}\right)  =-\ln\left(
\sum_{n=1}^{q}p_{n}\right)  =-\ln\left(  \sum_{n\in\Sigma}p_{n}\right)  .$
Therefore, $\operatorname*{dom}(\ln f)^{\ast}=[\theta_{1},{}\infty\lbrack{}.$

Summing up the preceding computations we get
\[
h^{\ast}(u,v)=\left\{
\begin{array}
[c]{ll}%
u\ln u-u+u\cdot(\ln f)^{\ast}(v/u) & \text{if }v\geq\theta_{1}u>0\\
-\alpha v & \text{if }u=0\leq v,\\
\infty & \text{if }u\in\mathbb{R}_{-}^{\ast},\text{ or }v\in\mathbb{R}%
_{-}^{\ast},\text{ or }0\leq v<\theta_{1}u.
\end{array}
\right.
\]
It follows that
\[
\left\{  (u,v)\in\mathbb{R}_{+}^{\ast}\times\mathbb{R}_{+}^{\ast}\mid
v>\theta_{1}u\right\}  =\operatorname*{int}(\operatorname*{dom}h^{\ast
})\subset\operatorname*{dom}h^{\ast}=\left\{  (u,v)\in\mathbb{R}_{+}%
\times\mathbb{R}_{+}\mid v\geq\theta_{1}u\right\}  .
\]
Moreover,
\begin{equation}
\partial h(\operatorname*{int}(\operatorname*{dom}h))=\nabla h(\mathbb{R}%
\times I)=\left\{  (u,v)\mid u\in\mathbb{R}_{+}^{\ast},\ \theta_{1}%
u<v<\theta_{2}u\right\}  =:E, \label{r-10}%
\end{equation}
as a simple argument shows.

For each $(u,v)\in\mathbb{R}^{2}$, $S(u,v):=S_{(\sigma_{n})}(u,v)$ and
$H(u,v):=H_{(\sigma_{n})}^{(p_{n})}(u,v)$ are defined in (\ref{r-su}) and
(\ref{r-hu}), respectively. By \cite[Prop.\ 15 (i)]{ValZal:15} we have that
$h^{\ast}(u,v)\leq H(u,v)$ for all $(u,v)\in\operatorname*{dom}h^{\ast}.$

In the sequel we determine the set $A$ of those $(u,v)\in\operatorname*{dom}%
h^{\ast}$ such that
\begin{equation}
h^{\ast}(u,v)=\min\bigg\{{\sum}_{n\geq1}u_{n}\big(\ln\frac{u_{n}}{p_{n}%
}-1\big)\mid(u_{n})_{n\geq1}\in S(u,v)\bigg\}, \label{r-11}%
\end{equation}
and the set $B$ of those $(u,v)\in\operatorname*{dom}h^{\ast}$ such that
$h^{\ast}(u,v)=H(u,v)$; of course, $A\subset B.$

Under our working hypothesis $(A_{\sigma f})$, we have that $\eta_{1}%
=\theta_{1}>0$, $\eta_{2}=\infty$ and
\[
\operatorname*{dom}H=\operatorname*{dom}S=\{(0,0)\}\cup\left\{  (u,v)\in
\mathbb{R}^{2}\mid v\geq\theta_{1}u>0\right\}  .
\]
Moreover, $S(0,0)=\{(0)_{n\geq1}\}$, while for $u>0$ and $\Sigma
:=\{k\in\mathbb{N}^{\ast}\mid\sigma_{k}=\theta_{1}\}=\overline{1,q}$ with
$q\in\mathbb{N}^{\ast},$
\[
S(u,\theta_{1}u)=\big\{(u_{n})_{n\geq1}\subset\mathbb{R}_{+}\mid
u_{n}=0~\forall n\geq q+1,\ u=u_{1}+\ldots+u_{q}\big\}.
\]

It follows that $h^{\ast}(0,v)=-\alpha v<\infty=H(0,v)$, whence $(0,v)\notin
B$, for $v>0$, and (\ref{r-11}) holds for $(u,v)=(0,0)$, and so $(0,0)\in A.$
Using \cite[Prop.\ 15 (v)]{ValZal:15}, we obtain that (\ref{r-11}) holds for
all $(u,v)\in E$ (with $E$ defined in (\ref{r-10})), with attainment for
$u_{n}:=p_{n}e^{x+\sigma_{n}y}$ $(n\geq1)$, where $y:=\varphi^{-1}(v/u)$ and
$x:=\ln\left[  u/f(y)\right]  $; hence $E\subset A.$

It remains to analyze the case of those $(u,v)\in\mathbb{R}_{+}^{\ast}%
\times\mathbb{R}_{+}^{\ast}$ with $v/u=\theta_{1}$ or $v/u\geq\theta_{2}$.

Let first $v=\theta_{1}u>0$. By Lemma \ref{lem2}, $H(u,v)$ is attained for
$\overline{u}_{n}=p_{n}u/\sum_{k=1}^{q}p_{k}=p_{n}u/\rho_{q}$ if
$n\in\overline{1,q}$, $\overline{u}_{n}:=0$ if $n\geq q+1$, and so
\[
H(u,v)=\sum\nolimits_{n=1}^{q}\frac{p_{n}u}{\rho_{q}}\bigg(\ln\frac{u}%
{\rho_{q}}-1\bigg)=u\big(\ln u-1-\ln\rho_{q}\big).
\]
Since
\[
h^{\ast}(u,v)=u\ln u-u+u\cdot(\ln f)^{\ast}(\theta_{1})=u\ln u-u-u\ln\rho
_{q}=H(u,v),
\]
we have that (\ref{r-11}) holds with attainment for $(\overline{u}_{n}%
)_{n\geq1}$ mentioned above; in particular $(u,v)\in A$.

Assume now that $\theta_{2}<\infty$, and so $-\alpha\in\operatorname*{dom}f$,
$\gamma<\infty$, and $\theta_{2}=\gamma/f(-\alpha)$. Take now $v\geq\theta
_{2}u>0$, and assume that $(u,v)\in A$. Since $v\geq\theta_{2}u>0$, by
(\ref{r-mbf7}), we have that $(u,v)\in\partial h(x,-\alpha)$ with $x:=\ln
\frac{u}{f(-\alpha)}$. Since $(u,v)\in A$, there exists $(u_{n})_{n\geq
1}\subset\mathbb{R}_{+}$ such that $(u,v)=\sum_{n\geq1}(u_{n},\sigma_{n}%
u_{n})$ and $h^{\ast}(u,v)=\sum_{n\geq1}u_{n}(\ln\frac{u_{n}}{p_{n}}%
-1)=\sum_{n\geq1}h_{n}^{\ast}(u_{n},\sigma_{n}u_{n})$. Using \cite[Prop.\ 15
(iv)]{ValZal:15}, we obtain that $(u_{n},\sigma_{n}u_{n})\in\partial
h_{n}(x,-\alpha)=\left\{  (p_{n}e^{x-\sigma_{n}\alpha},p_{n}\sigma
_{n}e^{x-\sigma_{n}\alpha})\right\}  $, that is $u_{n}=p_{n}e^{x-\sigma
_{n}\alpha}=p_{n}ue^{-\sigma_{n}\alpha}/f(-\alpha)$ $(n\geq1)$. It follows
that $v=\sum_{n\geq1}\sigma_{n}u_{n}=\frac{u}{f(-\alpha)}\sum_{n\geq1}%
p_{n}\sigma_{n}e^{-\sigma_{n}\alpha}=\frac{u}{f(-\alpha)}\gamma=\theta_{2}u$.
Conversely, if $v=\theta_{2}u>0$ and setting again $x:=\ln\frac{u}{f(-\alpha
)}$, the calculus above shows that $(p_{n}e^{x-\sigma_{n}\alpha})_{n\geq1}\in
A.$

Take now $v>\theta_{2}u>0$; we claim that $(u,v)\in B$, and so $(u,v)\in
B\setminus A.$

Let us set $x:=\ln\frac{u}{f(-\alpha)}$; then%
\begin{equation}
h^{\ast}(u,v)=u\ln u-u+u\cdot(\ln f)^{\ast}(v/u)=u\ln u-u-\alpha v-u\ln
f(-\alpha)=(x-1)u-\alpha v. \label{r-23}%
\end{equation}

Take $\overline{n}>\max\Sigma$ $(=q)$. Using Lemma \ref{lem1} we have that for
$n\geq\overline{n}$, $\varphi_{n}$ [definied in (\ref{r-fin})] is an
increasing bijection from $\mathbb{R}$ to $]\eta_{n}^{1},\eta_{n}^{2}[{}%
={}]\theta_{1},\eta_{n}^{2}[{}\subset{}]\theta_{1},\infty\lbrack{}$.
Moreover,
\[
\lim_{n\rightarrow\infty}\varphi_{n}(0)=\lim_{n\rightarrow\infty}\frac
{\sum_{k=1}^{n}p_{k}\sigma_{k}}{\sum_{k=1}^{n}p_{k}}=\lim_{n\rightarrow\infty
}\frac{p_{n+1}\sigma_{n+1}}{p_{n+1}}=\lim_{n\rightarrow\infty}\sigma
_{n+1}=\infty>\frac{v}{u},
\]
and
\[
\lim_{n\rightarrow\infty}\varphi_{n}(-\alpha)=\lim_{n\rightarrow\infty}%
\frac{\sum_{k=1}^{n}p_{k}\sigma_{k}e^{-\sigma_{k}\alpha}}{\sum_{k=1}^{n}%
p_{k}e^{-\sigma_{k}\alpha}}=\frac{\sum_{k\geq1}p_{k}\sigma_{k}e^{-\sigma
_{k}\alpha}}{\sum_{k\geq1}p_{k}e^{-\sigma_{k}\alpha}}=\frac{\gamma}%
{f(-\alpha)}=\theta_{2}<\frac{v}{u}.
\]
Increasing $\overline{n}$ if necessary, we may (and do) assume that
$\varphi_{n}(-\alpha)<v/u<\varphi_{n}(0)$ for $n\geq\overline{n}$. Hence, for
every $n\geq\overline{n}$ there exists a unique $\lambda_{n}\in{}%
]0,\alpha\lbrack$ with $\varphi_{n}(-\lambda_{n})=v/u$. Set $\upsilon_{n}%
:=\ln\big(u/\sum_{k=1}^{n}p_{k}e^{-\sigma_{k}\lambda_{n}}\big)$. Define
$u_{k}:=p_{k}e^{\upsilon_{n}-\sigma_{k}\lambda_{n}}$ for $k\in\overline{1,n}$
and $u_{k}:=0$ for $k>n$. Then
\[
\sum\nolimits_{k\geq1}u_{k}=\sum\nolimits_{k=1}^{n}p_{k}e^{\upsilon_{n}%
-\sigma_{k}\lambda_{n}}=u,\quad\sum\nolimits_{k\geq1}\sigma_{k}u_{k}%
=\sum\nolimits_{k=1}^{n}p_{k}\sigma_{k}e^{\upsilon_{n}-\sigma_{k}\lambda_{n}%
}=v.
\]
Because $\lambda_{n}<\alpha$ for $n\geq\overline{n}$, we get $\sum_{k=1}%
^{n}p_{k}e^{-\sigma_{k}\lambda_{n}}\geq\sum_{k=1}^{n}p_{k}e^{-\sigma_{k}%
\alpha}$, whence%
\[
\limsup_{n\rightarrow\infty}\upsilon_{n}\leq\limsup_{n\rightarrow\infty}%
\ln\frac{u}{\sum_{k=1}^{n}p_{k}e^{-\sigma_{k}\alpha}}=\ln\frac{u}{\sum
_{k\geq1}p_{k}e^{-\sigma_{k}\alpha}}=\ln\frac{u}{f(-\alpha)}=x.
\]
Moreover,%
\begin{align}
h^{\ast}(u,v)  &  \leq H(u,v)\leq\sum\nolimits_{k\geq1}u_{k}\bigg(\ln
\frac{u_{k}}{p_{k}}-1\bigg)=\sum\nolimits_{k=1}^{n}u_{k}(\upsilon_{n}%
-\sigma_{k}\lambda_{n}-1)\nonumber\\
&  =(\upsilon_{n}-1)\sum\nolimits_{k=1}^{n}u_{k}-\lambda_{n}\sum
\nolimits_{k=1}^{n}\sigma_{k}u_{k}=(\upsilon_{n}-1)u-\lambda_{n}v \label{r-22}%
\end{align}

Assume that $\liminf_{n\rightarrow\infty}\lambda_{n}<\alpha$. Then, for some
$\mu\in{}]0,\alpha\lbrack$ and some subsequence $(\lambda_{n_{m}})_{m\geq1}$
we have that $\lambda_{n_{m}}\leq\mu$ for every $m\geq1$. Then $\sum
_{k=1}^{n_{m}}p_{k}e^{-\sigma_{k}\lambda_{n_{m}}}\geq\sum_{k=1}^{n_{m}}%
p_{k}e^{-\sigma_{k}\mu}$, whence
\[
\liminf_{m\rightarrow\infty}\sum\nolimits_{k=1}^{n_{m}}p_{k}e^{-\sigma
_{k}\lambda_{n_{m}}}\geq\lim_{m\rightarrow\infty}\sum\nolimits_{k=1}^{n_{m}%
}p_{k}e^{-\sigma_{k}\mu}=\sum\nolimits_{k\geq1}p_{k}e^{-\sigma_{k}\mu}=\infty
\]
because $-\mu\notin\operatorname*{dom}f$. Since $\upsilon_{n_{m}}=\ln
u-\ln\sum_{k=1}^{n_{m}}p_{k}e^{-\sigma_{k}\lambda_{n_{m}}}$, we get
$\lim_{m\rightarrow\infty}\upsilon_{n_{m}}=-\infty$. Replacing $n$ by $n_{m}$
in (\ref{r-22}), then passing to the limit for $m\rightarrow\infty$, we get
the contradiction $-\infty<h^{\ast}(u,v)\leq u\cdot(-{}\infty)=-\infty$. Hence
$\lim\lambda_{n}=\alpha$ and $\limsup_{n\rightarrow\infty}\upsilon_{n}\leq x.$
Passing to $\limsup$ in (\ref{r-22}) for $n\rightarrow\infty$ we get
$H(u,v)\leq(x-1)u-\alpha v$, and so $h^{\ast}(u,v)=H(u,v).$

\medskip

Summing up the above discussion we get the next result.

\begin{theorem}
\label{t-mbe}Let $(p_{n})_{n\geq1}\subset\lbrack1,{}\infty\lbrack{},$
$(\sigma_{n})_{n\geq1}\subset\mathbb{R}_{+}^{\ast}$ with $\sigma
_{n}\rightarrow\infty$, and $h_{n}:\mathbb{R}^{2}\rightarrow\mathbb{R}$ be
defined by $h_{n}(x,y):=p_{n}e^{x+\sigma_{n}y}$ for $n\geq1$ and
$x,y\in\mathbb{R}$; set $h=\sum_{n\geq1}h_{n}$. Assume that
$\operatorname*{dom}h\neq\emptyset$. Clearly, $h$, $h_{n}$ $(n\geq1)$ are
convex and
\[
h(x,y)=e^{x}\sum\nolimits_{n\geq1}p_{n}e^{\sigma_{n}y}=e^{x}f(y)\quad
\forall(x,y)\in\mathbb{R}^{2},
\]
where $f$ is defined in (\ref{r-f}). Since $\operatorname*{dom}h=\mathbb{R}%
\times\operatorname*{dom}f\neq\emptyset$, using Proposition \ref{prop1}, we
have that $I:={}]\!-\infty,-\alpha\lbrack{}\subset\operatorname*{dom}%
f\subset\operatorname*{cl}I$ for some $\alpha\in\mathbb{R}_{+}$. It follows
that $\operatorname*{int}(\operatorname*{dom}h)=\mathbb{R}\times I\subset
\cap_{n\geq1}\operatorname*{dom}h_{n}=\mathbb{R}^{2}.$

\emph{(i)} We have that $h$ is differentiable on $\operatorname*{int}%
(\operatorname*{dom}h)$ and
\[
\partial h(\operatorname*{int}(\operatorname*{dom}h))=\nabla h(\mathbb{R}%
\times I)=\left\{  (u,v)\in\mathbb{R}^{2}\mid u\in\mathbb{R}_{+}^{\ast
},\ \theta_{1}u<v<\theta_{2}u\right\}  ,
\]
where $\theta_{1}:=\min\{\sigma_{n}\mid n\geq1\}$ and $\theta_{2}%
:=\lim_{y\uparrow-\alpha}f^{\prime}(y)/f(y)\in{}]\theta_{1},\infty];$
$\theta_{2}<\infty$ iff $-\alpha\in\operatorname*{dom}f$ and $\gamma
:=f_{-}^{\prime}(-\alpha)=\sum_{n\geq1}p_{n}\sigma_{n}e^{-\sigma_{n}\alpha
}<\infty$. Moreover, if $-\alpha\in\operatorname*{dom}f$ and $\gamma<\infty$
then
\[
e^{x}(f(-\alpha),\gamma)=\sum\nolimits_{n\geq1}\nabla h_{n}(x,-\alpha
)\in\partial h(x,-\alpha)=\{e^{x}f(-\alpha)\}\times\lbrack e^{x}\gamma
,{}\infty\lbrack{}.
\]

\emph{(ii)} The function $\varphi:I\rightarrow{}]\theta_{1},\theta_{2}[,$
$\varphi(y):=f^{\prime}(y)/f(y)$, is bijective (and increasing), $\ln
f:\mathbb{R}\rightarrow\overline{\mathbb{R}}$ is convex (even strictly convex
and increasing on its domain), and%
\[
(\ln f)^{\ast}\left(  w\right)  =\left\{
\begin{array}
[c]{ll}%
\infty & \text{if }w<\theta_{1},\\
-\ln\sum_{n\in\Sigma}p_{n} & \text{if }w=\theta_{1},\\
w\varphi^{-1}(w)-\ln\left[  f(\varphi^{-1}(w))\right]  & \text{if }\theta
_{1}<w<\theta_{2,}\\
-\alpha w-\ln\left[  f(-\alpha)\right]  & \text{if }\theta_{2}\leq w,
\end{array}
\right.
\]
where $\Sigma:=\{n\in\mathbb{N}^{\ast}\mid\sigma_{n}=\theta_{1}\}$. Moreover,
$\operatorname*{dom}h^{\ast}=\big\{(u,v)\in\mathbb{R}^{2}\mid v\geq\theta
_{1}u\geq0\big\}$ and
\[
h^{\ast}(u,v)=\left\{
\begin{array}
[c]{ll}%
u\ln u-u+u\cdot(\ln f)^{\ast}(v/u) & \text{if }v\geq\theta_{1}u>0,\\
-\alpha v & \text{if }u=0\leq v.
\end{array}
\right.
\]

\emph{(iii)} Take $(u,v)\in\operatorname*{dom}h^{\ast}$. Then
\[
h^{\ast}(u,v)=\min\bigg\{\sum\nolimits_{n\geq1}u_{n}(\ln\frac{u_{n}}{p_{n}%
}-1)\mid(u_{n})_{n\geq1}\in S(u,v)\bigg\}=H(u,v)
\]
iff $(u,v)\in A:=\{(0,0)\}\cup\{(u,v)\in\mathbb{R}_{+}^{\ast}\times
\mathbb{R}_{+}^{\ast}\mid\theta_{1}u\leq v\leq\theta_{2}u\}$, where
$S(u,v):=S_{(\sigma_{n})}(u,v)$ is defined in (\ref{r-su}) and $H:=H_{(\sigma
_{n}),E_{MB}}^{(p_{n})}$ is defined in (\ref{r-hu}). More precisely, for
$(u,v)\in A$ the minimum is attained at a unique sequence $(\overline{u}%
_{n})_{n\geq1}\in S(u,v)$, as follows: \emph{(a)} $(\overline{u}_{n})_{n\geq
1}$ $=(0)_{n\geq1}$ if $(u,v)=(0,0)$; \emph{(b)} $\overline{u}_{n}%
:=p_{n}u/\sum_{k\in\Sigma}p_{k}$ if $n\in\Sigma$, $\overline{u}_{n}:=0$ if
$n\in\mathbb{N}^{\ast}\setminus\Sigma$ provided $u\in\mathbb{R}_{+}^{\ast}$
and $v=\theta_{1}u$; \emph{(c)} $(\overline{u}_{n})_{n\geq1}=(p_{n}%
e^{x+\sigma_{n}y})_{n\geq1}$ if $u\in\mathbb{R}_{+}^{\ast}$ and $\theta
_{1}u<v<\theta_{2}u$, where $y:=\varphi^{-1}(v/u)$ and $x:=\ln\left[
u/f(y)\right]  $; \emph{(d)} $(\overline{u}_{n})_{n\geq1}=(p_{n}%
e^{x-\sigma_{n}\alpha})_{n\geq1}$ if $\theta_{2}<\infty$, $u\in\mathbb{R}%
_{+}^{\ast}$ and $v=\theta_{2}u$, where $x:=\ln\left[  u/f(-\alpha)\right]  .$

Moreover, $S(0,v)=\emptyset$ if $v\in\mathbb{R}_{+}^{\ast}$, and $h^{\ast
}(u,v)=H(u,v)$ whenever $0<\theta_{2}u<v$ (for $\theta_{2}<\infty$).
\end{theorem}

\begin{corollary}
Consider the sequences $(p_{n})_{n\geq1}\subset\lbrack1,{}\infty\lbrack{},$
$(\sigma_{n})_{n\geq1}\subset\mathbb{R}$ and let $H:=H_{(\sigma_{n})}%
^{(p_{n})}:=H_{(\sigma_{n}),E_{MB}}^{(p_{n})}$. Then $H^{\ast}(x,y)=\sum
_{n\geq1}p_{n}e^{x+\sigma_{n}y}=e^{x}\sum_{n\geq1}p_{n}e^{\sigma_{n}y}$ for
every $(x,y)\in\mathbb{R}^{2}.$
\end{corollary}

Proof. Assume first that the series $\sum_{n\geq1}p_{n}e^{\sigma_{n}y}$ is
divergent for each $y\in\mathbb{R}$. Then $h(x,y):=h_{(\sigma_{n}),E_{MB}%
}^{(p_{n})}(x,y):=\sum_{n\geq1}p_{n}e^{x+\sigma_{n}y}=\infty$ for all
$(x,y)\in\mathbb{R}^{2}$. Using Proposition \ref{p2} we have that $H$ takes
the value $-\infty$, and so $H^{\ast}(x,y)=\infty=h(x,y)$ for every
$(x,y)\in\mathbb{R}^{2}.$

Assume now that the series $\sum_{n\geq1}p_{n}e^{\sigma_{n}y}$ is convergent
for some $y\in\mathbb{R}$. By Proposition \ref{prop1} (i) we have that
$\sigma_{n}\rightarrow\infty$ or $\sigma_{n}\rightarrow-\infty.$

In the first case, if $\eta_{1}:=\theta_{1}:=\min_{n\geq1}\sigma_{n}>0$, using
Theorem \ref{t-mbe} we have that $\operatorname*{dom}H\subset
\operatorname*{dom}h^{\ast}=\left(  \{0\}\times\mathbb{R}_{+}^{\ast}\right)
\cup\operatorname*{dom}H$, $h^{\ast}\leq H$, and $h^{\ast}(u,v)=H(u,v)$ for
all $(u,v)\in\operatorname*{dom}H\supset\operatorname*{int}%
(\operatorname*{dom}h^{\ast})=\operatorname*{int}(\operatorname*{dom}H).$
Because for a convex function $f:E\rightarrow\overline{\mathbb{R}}$ with
$D:=\operatorname*{int}(\operatorname*{dom}f)\neq\emptyset$ one has $f^{\ast
}=(f+\iota_{D})^{\ast}$, it follows that $h=(h^{\ast})^{\ast}=H^{\ast}$. If
$\eta_{1}\leq0$, take $\sigma_{n}^{\prime}:=\sigma_{n}+a$ $(n\geq1)$ with
$a>-\eta_{1}$. Then $\left(  H_{(\sigma_{n}^{\prime})}^{(p_{n})}\right)
^{\ast}=h_{(\sigma_{n}^{\prime})}^{(p_{n})}$. But $h_{(\sigma_{n})}^{(p_{n}%
)}(x,y)=h_{(\sigma_{n}^{\prime})}^{(p_{n})}(x-ay,y)$ for $(x,y)\in
\mathbb{R}^{2}$ and $H_{(\sigma_{n})}^{(p_{n})}(u,v)=H_{(\sigma_{n}^{\prime}%
)}^{(p_{n})}(u,au+v)$ for $(u,v)\in\mathbb{R}^{2}$, whence
\[
\left(  H_{(\sigma_{n})}^{(p_{n})}\right)  ^{\ast}(x,y)=\left(  H_{(\sigma
_{n}^{\prime})}^{(p_{n})}\right)  ^{\ast}(x-ay,y)=h_{(\sigma_{n}^{\prime}%
)}^{(p_{n})}(x-ay,y)=h_{(\sigma_{n})}^{(p_{n})}(x,y)\quad\forall
(x,y)\in\mathbb{R}^{2}.
\]

If $\sigma_{n}\rightarrow-\infty$ take $\sigma_{n}^{\prime}:=-\sigma_{n}$
$(n\geq1)$. A similar argument as above shows that $\left(  H_{(\sigma_{n}%
)}^{(p_{n})}\right)  ^{\ast}=h_{(\sigma_{n})}^{(p_{n})}$. The proof is
complete.$\quad\square$

\section{Conclusions}

The Entropy Minimization Problem (EMP) is considered in Statistical Mechanics
and Statistical Physics for $W$ one of the functions $E_{MB,}$ $E_{BE}$,
$E_{FD}$. In general one obtains the optimal solutions using the Lagrange
multipliers method (LMM), method used by us in the proofs of Lemmas
\ref{lem2}, \ref{lem3a} and \ref{lem3b}. When the number of variables is
infinite this method can not be generally used because the function to be
minimized is not differentiable and the linear restrictions are not provided
by continuous (linear) operators (in this sense see the recent survey paper
\cite{Bor:12}). Even more, although the solutions found using LMM are indeed
solutions of the EMP, LMM does not provide always the solutions even in the
case of a finite numbers of variables as seen in Lemma \ref{lem3a} (iii).
Observe that in the works on Statistical Mechanics nothing is said about the
value of $(EMP)_{u,v}$ when the problem has not optimal solutions, and, of
course, if this value could be $-\infty$ or not; maybe this is not interesting
in Physics.

In the present paper, for $W=E_{MB}$, that is the Maxwell--Boltzmann entropy,
a complete study of the EMP is realized (when $p_{n}\geq1$ for $n\geq1$). More precisely,

-- the set of those $(u,v)\in\mathbb{R}^{2}$ for which $(EMP)_{u,v}$ has
feasible solutions is described (see Proposition \ref{p1});

-- it is shown that $H$ (the value function of the EMP) takes the value
$-\infty$ if and only if the series $\sum_{n\geq1}p_{n}e^{x+\sigma_{n}y}$ is
divergent for all $(x,y)\in\mathbb{R}^{2}$ (see Proposition \ref{p2});

-- when $\sum_{n\geq1}p_{n}e^{x+\sigma_{n}y}$ is convergent for some
$(x,y)\in\mathbb{R}^{2}$, it is confirmed that the solution found using LMM in
a formal way is indeed a solution of problem $(EMP)_{u,v};$ however, it is
shown that either there are situations in which $(EMP)_{u,v}$ has optimal
solutions not found using LMM, or there are situations in which $(EMP)_{u,v}$
has finite values but not optimal solutions (see Theorem \ref{t-mbe}).

\medskip

\textbf{Acknowledgement.} We thank Prof.\ M. Durea for his remarks on a
previous version of the manuscript.

\end{document}